\documentclass[sigplan,twocolumn,nonacm]{acmart}
\AtBeginDocument{%
  }

\setcopyright{none}
\acmConference{}{}{}
\renewcommand\footnotetextcopyrightpermission[1]{}
\usepackage[most]{tcolorbox}
\usepackage{algorithm}
\usepackage{algorithmic}
\usepackage{enumitem}
\usepackage{placeins}
\usepackage{subcaption}

\makeatletter
\newcommand{\removelatexerror}{\let\@latex@error\@gobble}
\makeatother

\newcommand{\eat}[1]{}

\newcommand{\refsec}[1]{Section~\ref{sec:#1}}

\newcommand{\stitle}[1]{\vspace{1ex} \noindent{{\bf #1}}}

\usepackage{xspace}

\newcommand{\name}[0]{SSV\xspace}

\begin{document}

%%
%% The "title" command has an optional parameter,
%% allowing the author to define a "short title" to be used in page headers.
\title{\name: Sparse Speculative Verification for Efficient LLM Inference}

%%
%% The "author" command and its associated commands are used to define
%% the authors and their affiliations.
%% Of note is the shared affiliation of the first two authors, and the
%% "authornote" and "authornotemark" commands
%% used to denote shared contribution to the research.
\author{Zhibin Wang}
% \authornote{Both authors contributed equally to this research.}
% \email{wzbwangzhibin@gmail.com}
% \orcid{1234-5678-9012}
\affiliation{%
  \institution{State Key Laboratory for Novel Software Technology}
  \institution{Nanjing University}
  \city{Nanjing}
  \country{China}
}

\author{Ziyu Zhong}
% \email{ziyuzhong@smail.nju.edu.cn}
\affiliation{%
  \institution{State Key Laboratory for Novel Software Technology}
  \institution{Nanjing University}
  \city{Nanjing}
  \country{China}
}

\author{Nuo Shen}
% \email{snnbyyds@gmail.com}
\affiliation{%
  \institution{State Key Laboratory for Novel Software Technology}
  \institution{Nanjing University}
  \city{Nanjing}
  \country{China}
}

\author{Yuhang Zhou}
% \email{yuhangzhou@smail.nju.edu.cn}
\affiliation{%
  \institution{State Key Laboratory for Novel Software Technology}
  \institution{Nanjing University}
  \city{Nanjing}
  \country{China}
}

\author{Rong Gu}
% \email{gurong@nju.edu.cn}
\affiliation{%
  \institution{State Key Laboratory for Novel Software Technology}
  \institution{Nanjing University}
  \city{Nanjing}
  \country{China}
}

\author{Sheng Zhong}
% \email{sheng.zhong@gmail.com}
\affiliation{%
  \institution{State Key Laboratory for Novel Software Technology}
  \institution{Nanjing University}
  \city{Nanjing}
  \country{China}
}

%%
%% By default, the full list of authors will be used in the page
%% headers. Often, this list is too long, and will overlap
%% other information printed in the page headers. This command allows
%% the author to define a more concise list
%% of authors' names for this purpose.
\renewcommand{\shortauthors}{Wang et al.}

%%
%% The abstract is a short summary of the work to be presented in the
%% article.
\begin{abstract}
  Speculative decoding and dynamic sparse attention are two complementary approaches for accelerating long-context LLM inference: the former amortizes target-model execution across multiple verifier queries, while the latter reduces each query's KV-cache working set. Directly combining them, however, exposes a structural mismatch: speculative verification relies on cross-query commonality, whereas dynamic sparse attention assigns query-specific sparse layouts. This mismatch limits KV-block reuse, amplifies NSA's branch-wise overheads, and makes verification strategy selection input- and regime-dependent. We present \name, a sparse speculative-verification framework that turns dynamic sparse attention into a verification-oriented workload. \name combines overlap-aware grouped-query execution, refresh/reuse-based NSA kernel fusion, and profile-guided prompt-adaptive orchestration to improve cross-query reuse, reduce selected-index and branch-fusion overheads, and select effective draft-verification strategies under user-specified precision classes.
  Experiments on NVIDIA H100 GPUs show that \name achieves up to $3.49\times$ end-to-end throughput over autoregressive NSA decoding and up to $6.86\times$ kernel speedups for sparse speculative verification.
\end{abstract}

%%
%% The code below is generated by the tool at http://dl.acm.org/ccs.cfm.
%% Please copy and paste the code instead of the example below.
%%
% \begin{CCSXML}
%   <ccs2012>
%   <concept>
%   <concept_id>00000000.0000000.0000000</concept_id>
%   <concept_desc>Do Not Use This Code, Generate the Correct Terms for Your Paper</concept_desc>
%   <concept_significance>500</concept_significance>
%   </concept>
%   <concept>
%   <concept_id>00000000.00000000.00000000</concept_id>
%   <concept_desc>Do Not Use This Code, Generate the Correct Terms for Your Paper</concept_desc>
%   <concept_significance>300</concept_significance>
%   </concept>
%   <concept>
%   <concept_id>00000000.00000000.00000000</concept_id>
%   <concept_desc>Do Not Use This Code, Generate the Correct Terms for Your Paper</concept_desc>
%   <concept_significance>100</concept_significance>
%   </concept>
%   <concept>
%   <concept_id>00000000.00000000.00000000</concept_id>
%   <concept_desc>Do Not Use This Code, Generate the Correct Terms for Your Paper</concept_desc>
%   <concept_significance>100</concept_significance>
%   </concept>
%   </ccs2012>
% \end{CCSXML}

% \ccsdesc[500]{Do Not Use This Code~Generate the Correct Terms for Your Paper}
% \ccsdesc[300]{Do Not Use This Code~Generate the Correct Terms for Your Paper}
% \ccsdesc{Do Not Use This Code~Generate the Correct Terms for Your Paper}
% \ccsdesc[100]{Do Not Use This Code~Generate the Correct Terms for Your Paper}

%%
%% Keywords. The author(s) should pick words that accurately describe
%% the work being presented. Separate the keywords with commas.
%% A "teaser" image appears between the author and affiliation
%% information and the body of the document, and typically spans the
%% page.

% Received dates are omitted for anonymous review submissions.

%%
%% This command processes the author and affiliation and title 
%% information and builds the first part of the formatted document.
\maketitle

\section{Introduction}
\label{sec:introduction}

Attention~\cite{vaswani2017attention} is the core mechanism of large language models (LLMs) that enables them to capture long-range dependencies in text.
During autoregressive decoding, generating each new token requires loading the entire accumulated key-value (KV) cache from memory to compute attention, which results in a heavy memory bandwidth cost, especially as the increasingly demanding of long-context applications~\cite{dao2023flashattention, kwon2023efficient}. Typically, for Llama-3.1-8B-Instruct~\cite{meta2024llama318binstruct,grattafiori2024llama3} with a 64K context on an NVIDIA H100 PCIe GPU~\cite{nvidia_h100}, the attention computation accounts for 97.20\% of each decoding step.
To relieve the pressure of KV-cache access, two complementary optimizations have emerged: speculative decoding~\cite{leviathan2023speculative,chen2023speculative} mitigates the bandwidth bottleneck by increasing arithmetic intensity, while sparse attention~\cite{gupta2021topk} directly reduces the volume of accessed KV cache.

Speculative decoding shifts the target model's execution toward the compute-bound by enlarging the effective query batch size. Instead of invoking the target model for each newly generated token, a lightweight draft model proposes candidate continuations, and the target model verifies multiple candidates in parallel. In dense verification, these verifier queries access the same committed-prefix KV cache simultaneously, amortizing the expensive KV-cache loads across multiple candidates. Many frameworks further organize candidates into draft trees~\cite{miao2024specinfer,cai2024medusa,li2024eagle,li2025eagle} to improve the acceptance rate. However, tree-structured verification introduces complex tree masks and speculative path dependencies. As the context size increases, these irregularities may also increase the access cost of KV cache.

Sparse attention takes another approach of shrinking the per-query KV working set. Instead of performing attention over the full KV history, dynamically selected sparse attention~\cite{gupta2021topk,yuan2025native,liu2025deepseek,deshmukh2025kascade,gao2026hysparse} routes each query to a subset of relevant context blocks, reducing both computation and KV-cache access in long-context decoding.
Taking the Native Sparse Attention (NSA)~\cite{yuan2025native} as an example, it combines block-level dynamic selection, compressed long-range context, and dense local windows with hardware-friendly sparse execution. Specifically, NSA first compresses the KV cache into blocks and selects the Top-$n$ relevant blocks for each query, while also preserving a sliding window of recent tokens. These branches are fused with learned gates to preserve complementary context while reducing KV access.

Intuitively, the two directions are naturally complementary, but their direct integration exposes a fundamental mismatch:

\emph{The per-query selectivity of dynamic sparse attention conflicts with the cross-query commonality of speculative verification.}

This mismatch creates three practical challenges:
(1) \textit{Cross-query reuse is hidden behind query-specific routing.} We observe that nearby verifier queries often select overlapping KV blocks, but decoding-oriented sparse kernels process each query independently and reload these blocks, while prefill-oriented kernels may fetch blocks unused by any verifier query. A practical verifier must exploit this overlap without losing the option to preserve exact per-query selected-block semantics.
(2) \textit{NSA's branch structure fragments short verifier work.} The compression, selection, and sliding-window branches are normally orchestrated through separate paths~\cite{yuan2025native}. During verification, the query batch is larger than single-token decoding but still too small to amortize repeated kernel launches and intermediate materialization.
(3) \textit{The best verification strategy is input-dependent.} Sparse verification changes how draft length, tree traversal, selected-block overlap, and refresh/reuse scheduling interact. As a result, a fixed speculative-decoding configuration can be suboptimal when acceptance behavior or sparse-kernel performance shifts across prompts and context-length regimes.

To address these challenges, we present \name, a sparse speculative-verification framework that integrates dynamic sparse attention into speculative decoding with three key mechanisms:

First, \name introduces an overlap-aware kernel design (\refsec{operator}) that flattens draft trees with selected traversal orders and groups up to $C$ adjacent verifier queries in one thread block. Each group can use an exact merged schedule that deduplicates overlapping selected blocks while preserving per-query NSA semantics, or an approximate shared-index layout that reuses one representative query's sparse layout for higher KV reuse.

Second, \name redesigns NSA kernel fusion around refresh and reuse layers (\refsec{kernel_fuse}). Refresh layers recompute selected indices and partially fuse the downstream selection and sliding-window branches, reducing branch-output materialization. Reuse layers inherit the nearest refresh indices, skip repeated index construction, and run a fully fused NSA path without intermediate branch writes.

Third, \name uses profile-guided prompt-adaptive orchestration (\refsec{planner}) to search complete strategy tuples covering draft-tree shape, traversal order, coarsening mode and factor, and refresh/reuse schedule. The planner ranks candidates offline for each context regime and precision class, then uses accepted-token observations to refine the selected strategy when runtime behavior deviates from the offline profile for that prompt.
In addition, \name allows users to specify precision classes that trade off acceptance rate and speed, supporting strict accuracy-preserving execution and faster controlled-approximation modes, while optimizing verification for the chosen setting.

In this paper, we make the following contributions:

\begin{itemize}[leftmargin=*, topsep=2pt, itemsep=0pt]
  \item We propose \name, a sparse speculative-verification framework that applies speculative decoding to dynamic sparse attention. \name adapts draft-tree verification to NSA-style sparse attention while preserving sparse-layout semantics and speculative accept/reject behavior.
  \item We introduce three optimizations that make sparse speculative verification efficient: overlap-aware kernel design for cross-query KV reuse, refresh/reuse-based NSA kernel fusion to reduce branch fragmentation and index-construction overhead, and profile-guided prompt-adaptive orchestration for complete strategy selection.
  \item We evaluate \name on NVIDIA H100 GPUs across isolated kernels, target verification, EAGLE-3 integration, and profile-guided planning, showing up to $6.86\times$ kernel speedup, $1.45\times$ verification-stage speedup, $3.49\times$ end-to-end throughput speedup, and $33.2\%$ higher accepted-token throughput from planning.
\end{itemize}

\section{Background}

\begin{figure*}[!t]
  \centering
  \includegraphics[width=\linewidth]{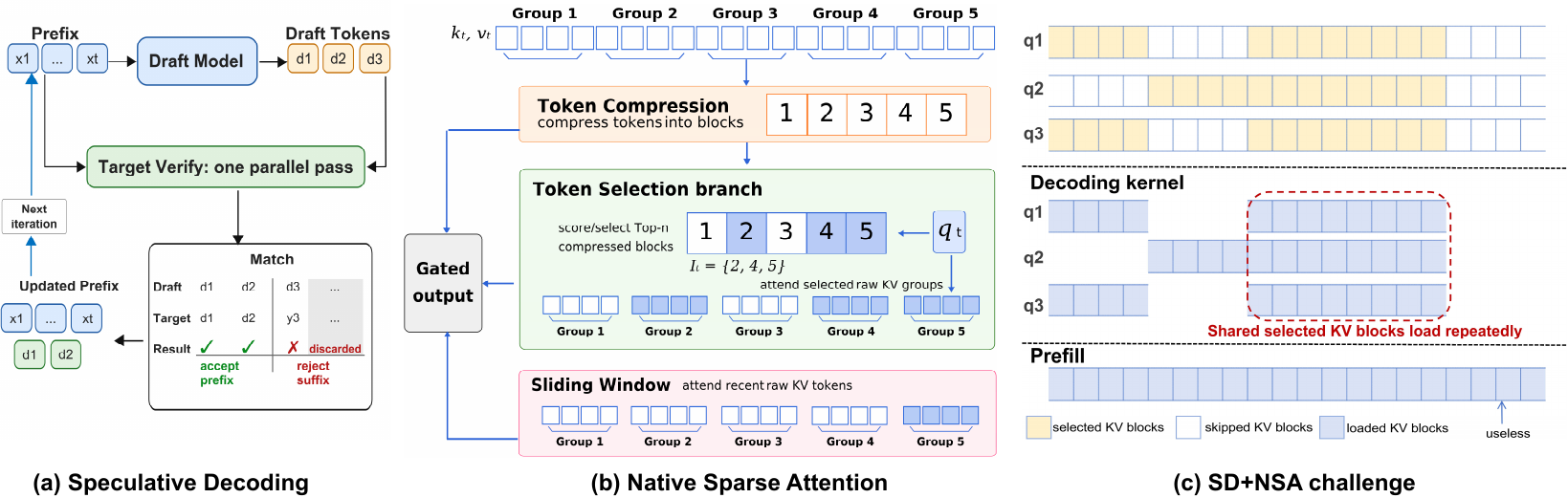}
  \caption{Background and mismatch when combining speculative decoding with sparse attention. }
  \label{fig:background_challenge}
\end{figure*}

\subsection{Speculative Decoding}
Speculative Decoding (SD)~\cite{leviathan2023speculative,chen2023speculative} accelerates LLM inference via a "draft-and-verify" workflow (Figure~\ref{fig:background_challenge}(a)). A lightweight draft model first proposes multiple candidate tokens, which are then verified concurrently by a larger target model. By verifying multiple candidate positions in a single forward pass, these queries share the prefix and access the same KV cache during attention. This amortizes memory access and execution overhead across candidate tokens, thereby increasing the effective query batch size and GPU utilization.

Recent research has further improved the acceptance rate by organizing candidates into trees~\cite{miao2024specinfer,cai2024medusa,li2024eagle,li2025eagle}. By allowing the draft model to expand multiple branches into a candidate tree, the target model can evaluate multiple potential paths concurrently, resulting in a higher acceptance rate but more complex verification. Specifically, tree-based speculation requires complex tree masks to track dependencies between draft tokens, which complicates attention execution. In addition, tree-based speculation increases the number of draft tokens per step, which requires more efficient kernel design to maintain the amortization benefits.

\subsection{Sparse Attention}
In long-context decoding, dense attention cost scales linearly with the KV-cache length. Sparse attention reduces this cost by restricting computation to a smaller working set of context tokens or blocks. Early methods rely on static patterns, such as sliding windows or predefined strided connections (e.g., Longformer~\cite{beltagy2020longformer}, BigBird~\cite{zaheer2020big}). However, static patterns struggle to capture the complex semantic relationships; dynamic methods (e.g., Top-$k$ attention~\cite{gupta2021topk}) have become the preferred standard. Recent systems often employ structured, hardware-aligned blockwise sparsity to better align KV-cache access with GPU execution~\cite{gao2026hysparse,yuan2025native}.
Among them, Native Sparse Attention (NSA)~\cite{yuan2025native} is a representative, and also the main focus of this paper.

As illustrated in Figure~\ref{fig:background_challenge}(b), NSA~\cite{yuan2025native} reduces KV access by fusing three complementary branches via learned gates:
(1) \textit{Compression branch:} Aggregates historical tokens into compressed blocks (length $l$, stride $d$) to capture broad context.
(2) \textit{Selection branch:} Uses compressed representations to route queries to the Top-$n$ relevant raw KV blocks (size $l'$). For a query $q_t$ at layer $j$, we denote these selected block indices as $I_t^{(j)}$.
(3) \textit{Sliding-window branch:} Retains dense access to the most recent $w$ tokens to preserve local semantics.
By operating at block granularity, the compression and selection branches ensure hardware-friendly, contiguous KV-cache access.

\stitle{Beyond NSA.}
Although \name utilizes NSA as its concrete backend, the architectural insights extend to other dynamic sparse-attention frameworks that decouple routing from execution. Extending to methods like DeepSeek Sparse Attention (DSA)~\cite{liu2025deepseek} is theoretically feasible but strictly conditioned. Specifically, transferring our approach requires: (1) explicitly exposed routing metadata, (2) the capability for query-level selection, and (3) sufficient cross-query index overlap. In the absence of these precise conditions, a backend-specific redesign is mandatory.

\subsection{Speculative Decoding Meets Sparse Attention}
The preceding text shows that SD and NSA optimize orthogonal dimensions of the attention bottleneck: SD amortizes target execution and dense KV-cache access across verifier queries, whereas NSA reduces the KV working set within each query. Intuitively, integrating them can provide a compounding acceleration for long-context inference. However, their direct combination exposes a structural mismatch:
\begin{tcolorbox}[
    enhanced,
    colback=gray!10,        % 浅灰背景
    colframe=black,         % 黑色边框
    boxrule=0.5pt,          % 边框粗细
    arc=1mm,                % 边角圆滑�?
    left=4pt, right=4pt, top=4pt, bottom=4pt, % 内边�?
    drop fuzzy shadow       % 淡淡的阴影增加立体感
  ]
  Speculative decoding depends on \textit{cross-query regularity}, whereas dynamic sparse attention introduces \textit{query-specific sparse layouts}.
\end{tcolorbox}
This structural mismatch arises because the two techniques extract efficiency from different forms of structure. Dense speculative verification benefits when many verifier queries remain aligned: they share the committed prefix, traverse the same KV-cache region, and therefore amortize target-model execution and KV-cache loads across the batch. Dynamic sparse attention, in contrast, derives efficiency from differentiating queries: each query independently selects a subset of relevant KV blocks, so its execution path is determined by its own routing result rather than by the batch-wide shared prefix.
In summary, combining them faces following challenges:

\stitle{Heavy redundant access.}
As shown in Figure~\ref{fig:background_challenge}(c), existing sparse-attention kernels can be classified into two categories: 1) \emph{Decoding-oriented kernels~\cite{yuan2025native,yan2025flash,liu2025deepseek,gao2026hysparse,lai2025flexprefill,deshmukh2025kascade}} that execute one query as well as load the corresponding KV blocks, and 2) \emph{Prefilling-oriented kernels~\cite{yuan2025native,liu2025deepseek,gao2026hysparse,deshmukh2025kascade}} that load the entire KV cache and for each token's KV, the query selecting the corresponding token are combined into one thread block. However, both categories fail to support the verification workload: employing decoding-oriented kernels leads to duplicated KV-block loads across queries, while prefilling-oriented kernels may load many KV blocks that are irrelevant to any query.

\stitle{Costly branch-wise kernel orchestration and intermediate data movement.}
NSA inherently evaluates compression, selection, and sliding-window branches and then combines their results. This structure is particularly awkward for speculative verification.
Unlike single-query decoding, verification involves multiple queries, which heavily amplifies branch-wise kernel launches and intermediate-result materialization. Unlike prefilling, the verifier batch is too small to amortize these fixed overheads, resulting in low arithmetic intensity per branch.
As a result, kernel launches, intermediate writes and reads, and cross-branch data movement account for a disproportionate share of latency. Full fusion is further limited because each branch maintains its own normalization state before gated aggregation.

\stitle{Input-dependent acceptance and varied kernel performance.}
Compared with speculative decoding with dense verification, the performance of sparse verification presents varied patterns related to the draft length. Dense attention uses a regular layout, whereas NSA changes the sparse execution plan as the draft length changes: selected-block sets, index-construction overhead, and cross-query layout overlap all depend on the generated draft positions. Even worse, we observe that the acceptance rate and kernel performance are both highly input-dependent (Table~\ref{tab:prompt_sensitivity}), which makes it difficult to select a single optimal configuration for all inputs.

\section{System Overview}
\label{sec:system_overview}

This section proposes \name to address the above challenges. We first summarize the design insights behind sparse speculative verification, and then show how \name integrates the corresponding mechanisms into the draft--verify--accept workflow.

\subsection{Design Insights}
\label{sec:key_observations}

\begin{figure}[t]
  \centering
  \includegraphics[width=\linewidth]{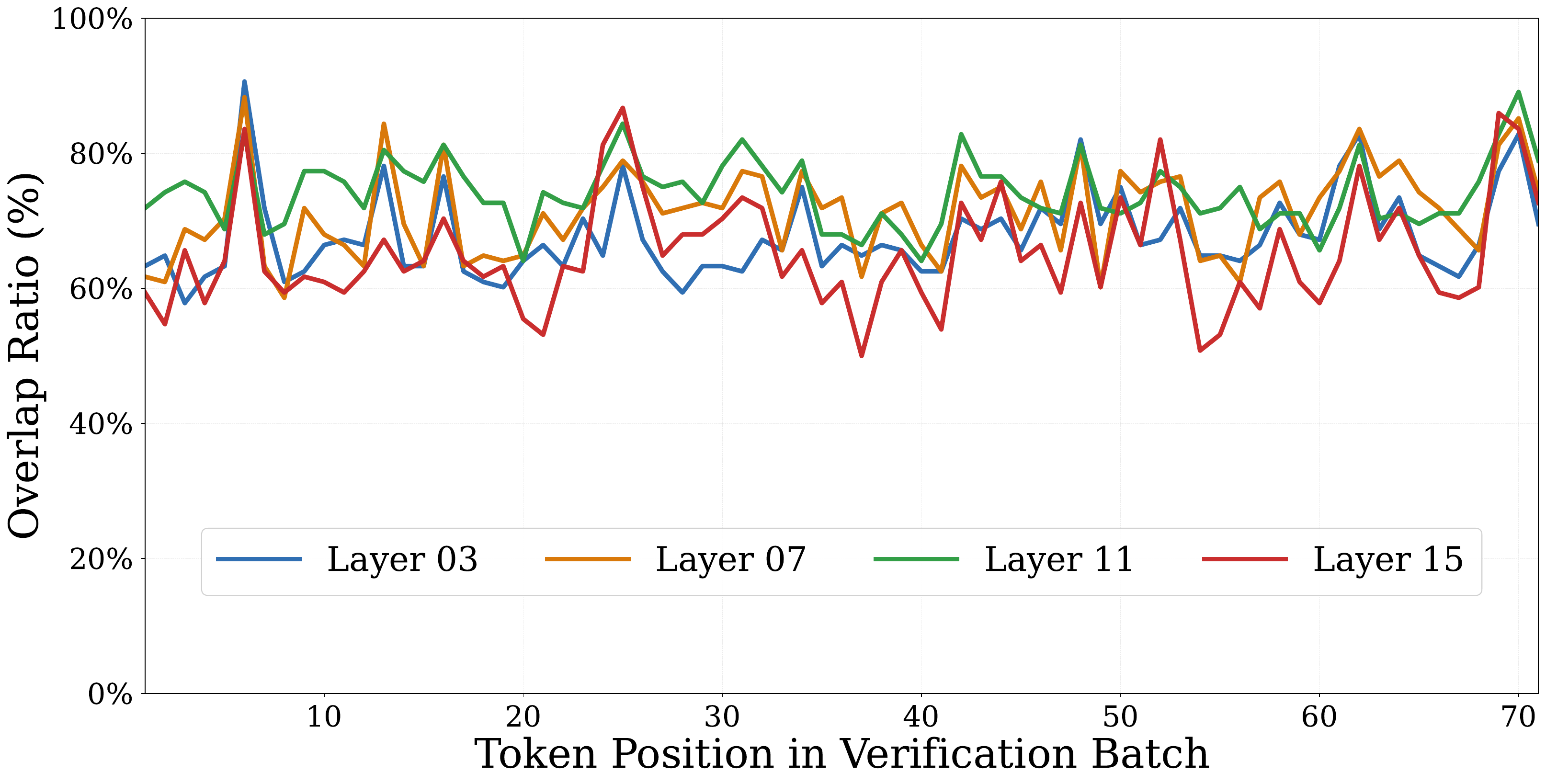}
  \caption{Selected-block overlap ratio between adjacent verifier queries (8K context).}
  % \caption{Selected-block overlap ratio between adjacent verifier queries under an 8K context. Across most layers and token positions, the overlap ratio stays around 50\%--90\%. Results are shown for the same NSA target model used in Section~\ref{sec:q1}.}
  \label{fig:adjacent_overlap_stats}
\end{figure}

\stitle{Insight 1: Nearby verifier queries have highly overlapping selected blocks.}
Tree-based speculative verification processes multiple draft positions in the same target-model pass. Although dynamic sparse attention gives each query its own selected-block set, verifier queries that are close in position often have semantically similar contexts and therefore select many of the same KV blocks. Our profiling confirms this trend: as shown in Figure~\ref{fig:adjacent_overlap_stats}, under an 8K context, the selected-block overlap ratio between adjacent verifier queries typically remains around 50\%--90\% across most layers. This overlap suggests grouping positionally close queries for joint execution, while the kernel can either preserve each query's own layout or use a faster approximate shared layout.

\stitle{Insight 2: Cross-layer stability enables fully fused execution.}
Selected-block layouts are often stable across nearby Transformer layers, consistent with prior observations~\cite{bai2026indexcache,deshmukh2025kascade,gao2026hysparse}; our profiling confirms the same trend during speculative verification (Table~\ref{tab:approx_quality}). \name therefore organizes execution into Refresh/Reuse layers: refresh layers recompute selected indices and retain the routing-aware partial-fusion path, whereas reuse layers inherit the latest layout from the nearest preceding refresh layer, bypass index derivation, and directly enter a fully fused kernel.

\stitle{Insight 3: Long decode horizons enable prompt-aware adaptive planning.}
Speculative decoding typically spans many verification rounds when generating a response, which gives the system a natural adaptation window for each prompt. Rather than committing to one fixed strategy for the entire generation, \name can start from a strong initial choice and refine it as prompt-specific acceptance behavior becomes observable. This observation motivates \name{}'s throughput-aware planner, which combines offline strategy preselection with low-overhead prompt-aware reselection during decoding.

\subsection{Overview}
\begin{figure}[t]
  \centering
  \includegraphics[width=\linewidth]{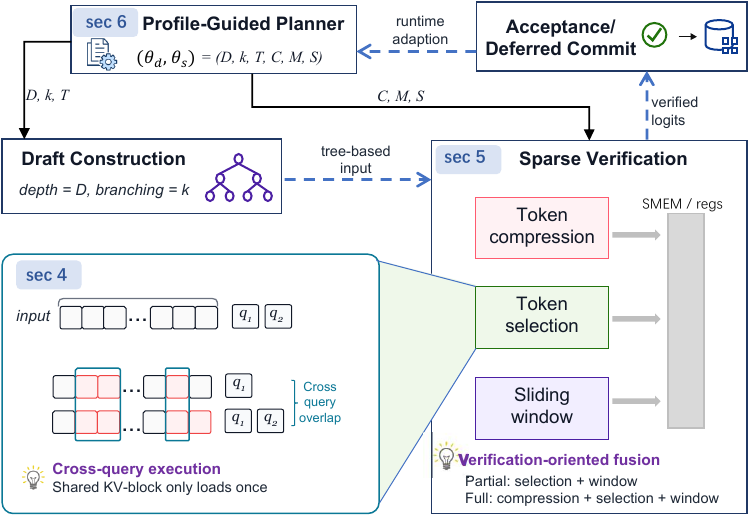}
  \caption{Overview of \name.}
  % \caption{Overview of \name's workflow. A profile-guided planner selects a joint strategy for draft construction and sparse verification. The verifier exploits overlap in selected block indices and executes NSA branches through fused paths: \texttt{F} layers fuse selection and window attention, while \texttt{S} layers reuse selected block indices to fuse all three branches through shared memory/register state. Verified logits are then handled by the acceptance/deferred-commit logic, which can feed runtime adaptation back to the planner.}
  \label{fig:workflow}
  \Description{}
\end{figure}

Figure~\ref{fig:workflow} illustrates how \name integrates these mechanisms into the standard draft-verify-accept loop. While the outer interface remains unchanged, \name fundamentally overhauls the verification internals. Rather than naively replacing dense attention with an off-the-shelf sparse kernel, \name treats verification as a jointly planned workload: the planner synergizes draft construction with sparse verification, while the verifier exploits cross-query index overlap and fused execution to maximize hardware efficiency.

\stitle{Overlap-aware cross-query execution (\refsec{operator}).}
The first optimization module targets redundant KV-block traffic inside sparse verification. \name groups nearby verifier queries according to the draft traversal order and executes them within a shared thread block, so overlapping selected blocks are loaded once and reused across queries. Depending on the deployment precision budget, this module either preserves per-query sparse layouts through an exact merged schedule or uses a faster approximate shared-index layout for highly overlapping query groups.

\stitle{Verification-oriented kernel fusion (\refsec{kernel_fuse}).}
The second module reduces the branch fragmentation introduced by NSA's compression, token-selection, and sliding-window paths. In refresh layers, \name keeps the routing-dependent compression path separate but fuses token selection, sliding-window attention, and gated aggregation into a downstream kernel. In reuse layers, stable selected-block indices remove the routing dependence, enabling a fully fused kernel that computes all NSA branches with on-chip intermediate state and writes back only the final output.

\stitle{Profile-guided adaptive planning (\refsec{planner}).}
The third module coordinates draft construction and sparse verification as one strategy-selection problem. An offline profiler ranks joint configurations over draft-tree shape, traversal order, cross-query coarsening, and refresh/reuse schedules for each context regime and precision class. At runtime, \name starts from the profiled best strategy and refines it during the early decoding window when prompt-specific acceptance or kernel behavior deviates from the profile.

\section{Overlap-Aware Kernel Design}
\label{sec:operator}

\begin{figure}[t]
    \centering
    \includegraphics[width=\linewidth]{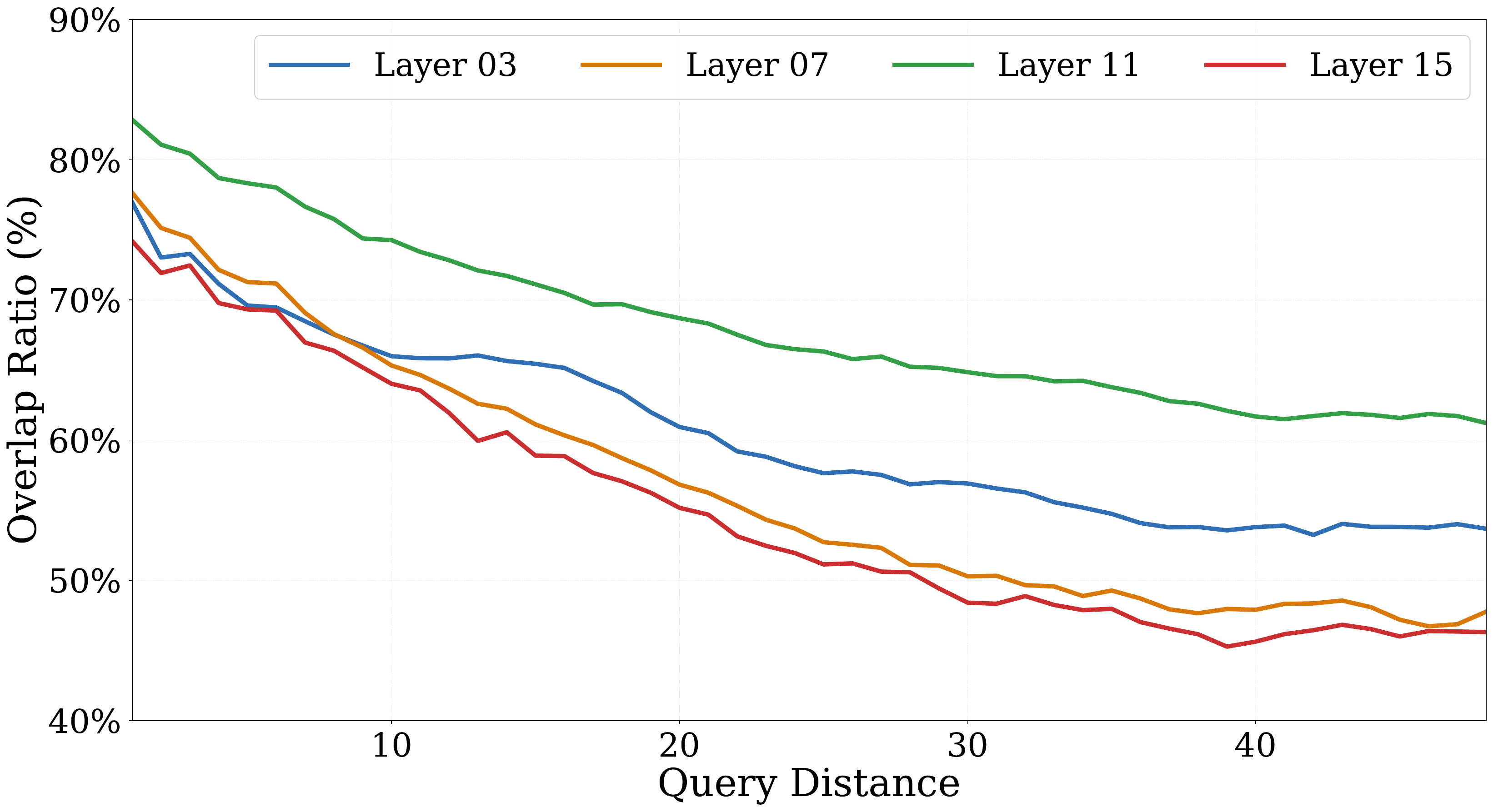}
    \caption{Selected-block overlap ratio versus absolute token-position distance $\Delta$ (8K context). The cross-query overlap consistently peaks at small $\Delta$ and decays as queries become further apart in the sequence.}
    % \caption{Selected-block overlap ratio versus token-position distance $\Delta$ under 8K context. Here $\Delta$ denotes the absolute sequence-position difference between two verifier queries, rather than their distance in the flattened verification batch. Across layers, the mean shared selected-block overlap ratio is highest at small $\Delta$ and generally decreases as $\Delta$ grows.}
    \label{fig:topk_idx_over_delta}
    \Description{}
\end{figure}

Insight~1 and Figure~\ref{fig:topk_idx_over_delta} show that verifier queries that are close in sequence position tend to select overlapping KV blocks, with overlap decreasing as token-position distance grows.
Exploiting this spatial locality, \name employs overlap-aware cross-query coarsening, grouping adjacent queries into a single thread block to amortize KV-block loads and scheduling overheads. We develop two variants of kernel execution: an \emph{exact} merged-schedule variant that preserves precise per-query selection semantics, and an \emph{approximate} shared-index variant that aggressively reuses a single selected-block set across the group, trading exact selection semantics for lower scheduling overhead.
% Motivated by this observation, \name organizes sparse verification around overlap-aware cross-query coarsening: it groups nearby queries into the same thread block so they can share KV-block loads and related scheduling work. Subsequently, we develop two variants of kernel execution: the exact merged-schedule variant preserves the original per-query selection semantics by constructing a merged schedule and applying row-wise masks, while the approximate shared-index variant reuses one query's selected-block set across the group, trading exact selection semantics for lower scheduling overhead and stronger KV-block reuse.

\subsection{Cross-Query Grouping}
\label{sec:cross_query_grouping}

% Before delving into the grouping design, we first consider two naive grouping strategies. \emph{Single-query grouping} assigns one query per thread block, which preserves the original sparse-attention semantics but fails to exploit cross-query overlap. \emph{All-query grouping} assigns all queries to the same thread block, which maximizes cross-query reuse but 1) suffers from heavy scheduling overhead (accuracy loss) due to the large query set and long token distance, and 2) can not fully exploit hardware ability due to limited parallelism.

\begin{figure}[t]
    \centering
    \includegraphics[width=\linewidth]{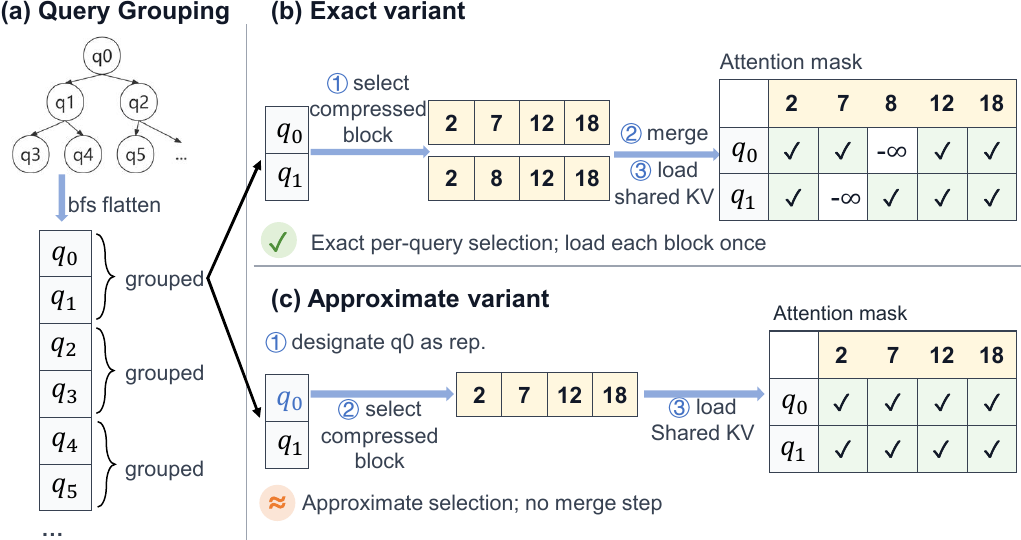}
    \caption{Overlap-aware kernel design.}
    \label{fig:thread_coarsening}
    \Description{}
\end{figure}

\stitle{Grouping design.}
To balance cross-query KV reuse and kernel scheduling overhead, \name avoids the extremes of single-query grouping (zero reuse) and all-query grouping (massive scheduling overhead and low parallelism). Instead, as shown in Figure~\ref{fig:thread_coarsening}(a), it employs a middle-ground strategy: partitioning verifier queries into thread blocks of up to $C$ adjacent queries.
This design balances the tradeoff between reuse and overhead by keeping the group size $C$ manageable.
% Therefore, as shown in Figure~\ref{fig:thread_coarsening}(a), \name uses a middle-ground strategy that groups up to $C$ adjacent queries into the same thread block. This design balances the tradeoff between reuse and overhead by exposing some cross-query overlap while keeping the group size manageable for scheduling and parallelism.

While grouping is trivial for a flat draft sequence, tree-based speculation introduces structural differences. Specifically, a tree node can be grouped either with its siblings (candidates for the same position that are likely synonymous) or with its parent and children (candidates across nearby positions with strong semantic proximity). In practice, \name reorganizes the draft tree into a flattened batch using two traversal orders~\cite{cormen2022introduction}: breadth-first search (BFS) to enforce sibling-based grouping, and depth-first search (DFS) to enforce parent/child-based grouping.

Because the optimal grouping strategy depends heavily on the specific verification setting and realized query ordering, no single tree-local rule is universally superior. Therefore, rather than using a static heuristic, \name delegates this choice to the throughput-aware planner (\refsec{planner}), which adaptively selects the best traversal order based on offline profiles and runtime states.
% The grouping is trivial when the draft is a flat sequence, while the mainstream tree-based draft makes the grouping design more nuanced. Specifically, for a node in the draft tree, it can be grouped with its siblings or parent/children. Both strategies have their own reasons: sibling nodes are candidates for same position, which are highly likely to be synonymous; parent/child nodes are candidates for nearby positions with semantic proximity.
% Our primary evaluation shows that the choice of grouping strategy depends on the actual verification setting and the realized verifier-query ordering, rather than admitting a single universally optimal tree-local rule. As discussed in \refsec{planner}, \name will adaptively select the grouping strategy according to the profile and runtime state.

% In practice, we find that two common traversal orders, breadth-first search (BFS) and depth-first search (DFS)~\cite{cormen2022introduction}, can be used to implement sibling-based and parent/child-based grouping, respectively. BFS places nodes with the same sequence position together, while DFS places nodes with nearby positions together. We reorganize the output of the draft model to support both traversal orders when materializing the draft tokens into a flattened batch.

Driven by the selected traversal order, \name partitions the flattened batch into groups of up to $C$ adjacent queries. Following the principle of thread coarsening~\cite{wen2026programming}, each group is processed in parallel by a single thread block. By sharing register tiles across query heads in grouped-query attention (GQA)~\cite{ainslie2023gqa}, this execution amortizes memory access and scheduling overheads without altering the logical tree masks or absolute positions. This grouped execution naturally extends to the sliding-window branch, which advances a shared window scan while enforcing per-query local masks.

After grouping verifier queries, the remaining question is how the kernel handles their non-identical selected-block sets. We will delve into two distinct execution variants.

\subsection{Exact Merged-Schedule Variant}
\label{sec:exact_coarsening}
The exact merged-schedule variant reduces memory traffic by jointly scheduling shared KV blocks without altering per-query selection semantics. As shown in Figure~\ref{fig:thread_coarsening}(b), the kernel achieves this through three steps:
(1) \textit{Independent Selection:} It computes the selected block-index set $I_t^{(j)}$ independently for each query in the group.
(2) \textit{On-chip Merging:} It concatenates these per-query indices, sorts them, and performs a linear scan to deduplicate identical entries. This creates a unified schedule containing the union of all selected blocks while tracking query-to-block ownership.
(3) \textit{Shared Loading and Masking:} Each unique KV block in the merged schedule is loaded from HBM exactly once and shared across the group. To preserve exact semantics, the kernel applies a row-wise mask during attention computation: logits are masked to $-\infty$ if the key block does not belong to a specific query's original selected set.

This design effectively amortizes the HBM load costs of overlapping blocks while remaining semantically equivalent to independent query execution.

\subsection{Approximate Shared-index Variant}
\label{sec:approximate_coarsening}
Prior work~\cite{matx2025sd_nsa} has shown that leveraging the same index for neighboring tokens barely affects model quality. Therefore, this variant aims to maximize reuse by forcing all $C$ queries in the group to share the same selected block set.
As shown in Figure~\ref{fig:thread_coarsening}(c), instead of computing per-query indices and merging them, the kernel simply: 1) designates a single representative query for the group (e.g., the one with the longest prefix); 2) computes the selected block-index set $I_t^{(j)}$ solely for this representative; and 3) applies this shared layout to compute attention for the entire group.

This strategy completely bypasses the sort-and-merge overhead and drastically reduces HBM traffic for KV-block loading. The approximation arises because non-representative queries may miss their individually preferred KV blocks. However, while the sparse-attention branch becomes approximate, the sliding-window branch remains entirely precise, enforcing strict per-row boundaries and causal masks using each query's individual absolute position.

\stitle{Accuracy evaluation.} To validate its quality impact, we conduct a model-integrity study on a 1B NSA target model~\cite{zen2025nsa1b,ssa_repo}. We evaluate the variant using lm-evaluation-harness~\cite{eval-harness} on PIQA~\cite{bisk2020piqa}, HellaSwag~\cite{zellers2019hellaswag}, ARC-Easy, and ARC-Challenge~\cite{clark2018think} (using 3-, 10-, and 25-shot, respectively). As shown in Table~\ref{tab:approx_quality}, the approximate variant ($C=4$) exhibits negligible accuracy degradation compared to the exact \name baseline. In practice, \name applies this aggressive coarsening only in regimes with sufficient concurrent queries (typically at larger draft length $\gamma$) where cross-query reuse heavily outweighs the approximation penalty.

\begin{table}[t]
    \centering
    \footnotesize
    \setlength{\tabcolsep}{3.6pt}
    \caption{Benchmark results (\%) on a 1B NSA model, evaluated on PIQA, HellaSwag, ARC-Easy, and ARC-Challenge. ``\name + reuse layers'' uses the training-free IndexCache-style~\cite{bai2026indexcache} refresh/reuse schedule \(\mathcal{S}=\{3,6,7,8,12,13,14,15\}\), and ``$C=4$'' denotes the approximate shared-index variant with grouping factor 4.}
    \label{tab:approx_quality}
    \begin{tabular}{@{}lcccc@{}}
        \toprule
        \shortstack[t]{\textbf{Benchmark}      \\\mbox{}} & \shortstack[t]{\textbf{\name}\\\mbox{}} & \shortstack[t]{\textbf{\name}\\\textbf{+ reuse}} & \shortstack[t]{\textbf{\name}\\\textbf{($C=4$)}} & \shortstack[t]{\textbf{\name + reuse}\\\textbf{($C=4$)}} \\
        \midrule
        PIQA / \%          & \shortstack{73.78 \\$\pm$ 1.03} & \shortstack{73.99\\$\pm$ 1.02} & \shortstack{73.88\\$\pm$ 1.02} & \shortstack{73.88\\$\pm$ 1.02} \\
        HellaSwag / \%     & \shortstack{58.37 \\$\pm$ 0.49} & \shortstack{58.55\\$\pm$ 0.49} & \shortstack{58.53\\$\pm$ 0.49} & \shortstack{58.47\\$\pm$ 0.49} \\
        ARC-Easy / \%      & \shortstack{69.49 \\$\pm$ 0.94} & \shortstack{69.57\\$\pm$ 0.94} & \shortstack{69.82\\$\pm$ 0.94} & \shortstack{69.82\\$\pm$ 0.94} \\
        ARC-Challenge / \% & \shortstack{36.95 \\$\pm$ 1.41} & \shortstack{37.12\\$\pm$ 1.41} & \shortstack{37.37\\$\pm$ 1.41} & \shortstack{37.29\\$\pm$ 1.41} \\
        \bottomrule
    \end{tabular}
\end{table}

% The approximation arises because queries other than the representative one may miss KV blocks they would have selected under independent execution, and may instead attend to blocks selected only by that representative query. Consequently, this variant no longer preserves the exact selected block set for each query, and may introduce output deviations, although the degradation is often small when neighboring queries already have strong overlap. Its quality impact is evaluated in Table~\ref{tab:approx_quality}, where we show that the resulting degradation remains limited in our model-integrity study. In practice, we apply the approximate variant of thread coarsening only in regimes where verification contains enough concurrent queries for cross-query reuse to dominate the cost of schedule construction and extra KV traffic, typically at larger $\gamma$ and stronger overlap.

\section{Kernel Fusion}
\label{sec:kernel_fuse}

\begin{figure}[tbp]
    \centering
    \includegraphics[width=\linewidth]{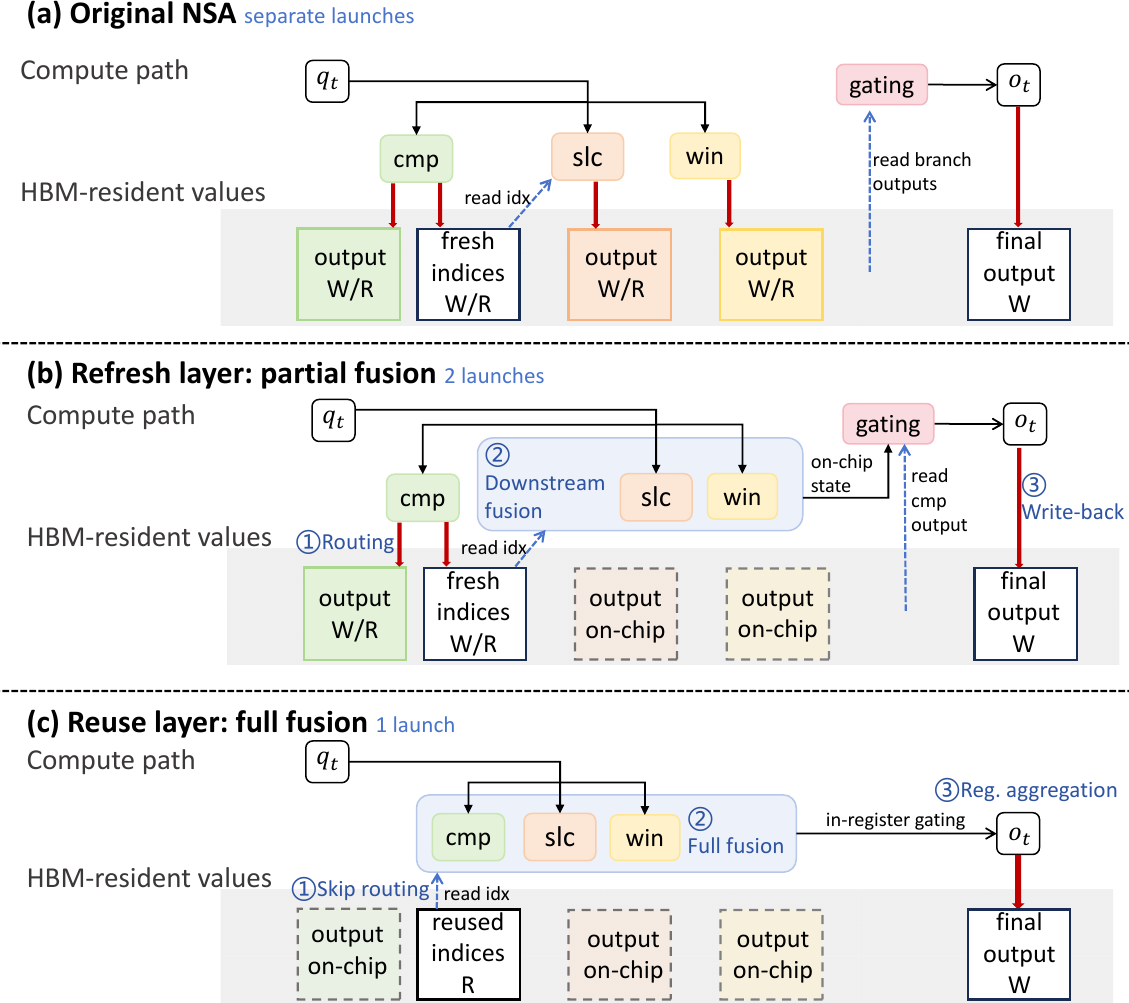}
    \caption{Demonstration of three fusion strategies.}
    % \caption{Verification-oriented NSA fusion with \emph{refresh}/\emph{reuse} index scheduling in \name.
    % The lower row reports HBM materialization (\texttt{W/R}: write-and-reread; \texttt{W}: final writeback; \texttt{R}: reused-index read; on-chip: not materialized).
    % Original NSA materializes the selected indices and all branch outputs; partial fusion removes separate \(O_{\mathrm{slc}}\) and \(O_{\mathrm{win}}\) materialization; full fusion reuses \(I_t^{(j_{\mathrm{prev}})}\), fuses all branches, and writes back only \(O_t\).
    % Common KV-cache reads are omitted.}
    \label{fig:kernel_fusion}
    \Description{}
\end{figure}

Figure~\ref{fig:kernel_fusion}(a) illustrates the branch-wise execution challenge of the original NSA. While evaluating branches through separate paths is tolerable in standard single-query decoding, it becomes a severe bottleneck in speculative verification. Because a verifier pass consists of many short, heterogeneous queries, repeatedly materializing intermediate indices and branch outputs to HBM for gated aggregation incurs disproportionate kernel-launch and memory-bound overheads.

To eliminate this fragmentation, \name employs a two-tiered kernel fusion strategy guided by cross-layer index stability (Insight~2)~\cite{bai2026indexcache}:
(1) \textit{Routing-aware partial fusion (Refresh layers):} For layers that must compute fresh indices, \name respects the strict data dependence between routing and selected attention. It isolates the compression/routing path while fusing the downstream token-selection and sliding-window branches.
(2) \textit{Index-reuse-enabled full fusion (Reuse layers):} For layers that inherit a stable selected-block layout from a preceding refresh layer, the data dependence is broken. \name bypasses index derivation entirely and deploys a fully fused kernel that computes all three NSA branches and their gated aggregation in a single launch.

\subsection{Routing-Aware Partial Fusion}
NSA leaves a natural partial-fusion opportunity: the token-selection and sliding-window branches both consume the same verifier query and can share one downstream kernel, provided their independent normalization states are preserved. However, the compression/routing path must remain isolated in refresh layers for two reasons. First, there is a strict \emph{data dependency}: selected attention cannot issue sparse KV loads until routing produces the Top-$n$ indices. Second, they exhibit \emph{mismatched hardware execution patterns}: routing is bound by importance scoring and Top-$n$ reduction, whereas downstream attention is optimized for sparse memory gathering and attention accumulation. Forcing them into a monolithic kernel would couple conflicting resource requirements and offset the fusion benefits.
% NSA leaves a natural partial-fusion opportunity after sparse routing. The token-selection and sliding-window branches both consume the same verifier query and run independent attention computations before gated aggregation. They can therefore share one downstream kernel as long as the kernel preserves each branch's normalization state.
% \wzb{refine the following sentences (sn do it).}\sn{OK. Besides data dependency, I emphasize different
%     hardware-efficient execution patterns here.}
% The compression/routing path is different: in refresh layers, it must first produce the selected indices that determine which raw KV blocks the selected attention will visit. Thus, selected attention depends on the routing result before it can issue sparse KV loads. More importantly, routing and downstream sparse attention favor different hardware-efficient execution patterns: routing is dominated by importance score computation and Top-$n$ block selection, whereas the downstream branches are optimized for sparse block gathering and attention accumulation. Forcing them into one monolithic kernel would couple these mismatched parallelization patterns and resource requirements, often reducing effective occupancy and offsetting the benefit of deeper fusion.

To resolve this, \name executes refresh layers using a two-launch strategy, as shown in Figure~\ref{fig:kernel_fusion}(b):
(1) \textit{Routing Launch:} \name first runs the compression/routing path to produce the compressed-branch output and the fresh selected indices.
(2) \textit{Downstream Fusion:} Once indices are available, \name launches a partially fused kernel for the token-selection and sliding-window branches. It reads the newly generated indices, computes both attention branches alongside their independent online softmax states, and performs gated aggregation directly in registers.
(3) \textit{Unified Write-back:} The kernel writes only the final aggregated layer output back to HBM, completely bypassing intermediate branch-output materialization.
% Figure~\ref{fig:kernel_fusion}(b) illustrates the resulting two-launch execution. (1) \name first runs the compression/routing path as a separate launch, producing the compressed-branch output and fresh selected indices. (2) After these indices are available, \name launches a partially fused downstream kernel for the token-selection and sliding-window branches; the kernel reads the selected indices, computes the two branches together with independent online softmax states, combines their gated contribution in registers, and avoids writing their branch outputs back to HBM separately. (3) The final aggregation writes back only the layer output. \wzb{mark workflow 1,2,3 in the figure and add 1,2,3 in the contex}\zzy{Added a third step to correspond with the figure}
This removes downstream branch fragmentation but still leaves selected-index derivation outside the fused kernel, motivating the index-reuse-enabled full-fusion path introduced next.

\subsection{Index-Reuse-Enabled Full Fusion}
Prior cross-layer index reuse studies observe that sparse indices are often stable across adjacent Transformer layers~\cite{bai2026indexcache,deshmukh2025kascade}. Therefore, for \emph{reuse layers}, the selected-block layout is inherited from the nearest preceding \emph{refresh layer}~\cite{bai2026indexcache,deshmukh2025kascade}. This inheritance provides a controlled approximation that bypasses layer-specific recomputation overhead while preserving concrete sparse evaluation. To initialize the stream, the first layer acts as a mandatory refresh layer.

As illustrated in Figure~\ref{fig:kernel_fusion}(c), \name exploits this by enabling a single-launch full-fusion path:
(1) \textit{Zero-Overhead Routing:} Because the sparse layout is already available, \name completely bypasses the separate routing launch.
(2) \textit{Monolithic Execution:} It invokes a fully fused kernel that reads the inherited indices, interprets them under the current draft-position masks and compressed-block visibility constraints, and computes the compression, token-selection, and sliding-window branches in one unified pass.
(3) \textit{In-Register Aggregation:} The kernel keeps all intermediate branch outputs on-chip, performs gated aggregation directly in registers, and writes back only the final layer output to HBM.
By transforming the expensive indices from a refresh layer into a reusable execution plan, this design drastically reduces launch overhead and avoids intermediate materialization.

\stitle{Quality impact.}
Because index reuse is a controlled approximation, we evaluate its impact on model integrity using a 1B NSA target model~\cite{zen2025nsa1b,ssa_repo}. We apply a IndexCache-style~\cite{bai2026indexcache} greedy calibration to determine the optimal schedule, yielding the reuse pattern \(\mathcal{S}=\{3,6,7,8,12,13,14,15\}\). Evaluated via lm-evaluation-harness~\cite{eval-harness} on PIQA, HellaSwag, ARC-Easy, and ARC-Challenge, both the standard index-reuse execution and its aggressive combination with approximate cross-query grouping ($C=4$) exhibit negligible accuracy degradation (Table~\ref{tab:approx_quality}). This confirms that our schedule preserves reasoning capabilities while maximizing hardware efficiency.

\FloatBarrier
\section{Profile-Guided Prompt-Adaptive Orchestration}
\label{sec:planner}

\begin{figure}[t]
    \centering
    \includegraphics[width=\linewidth]{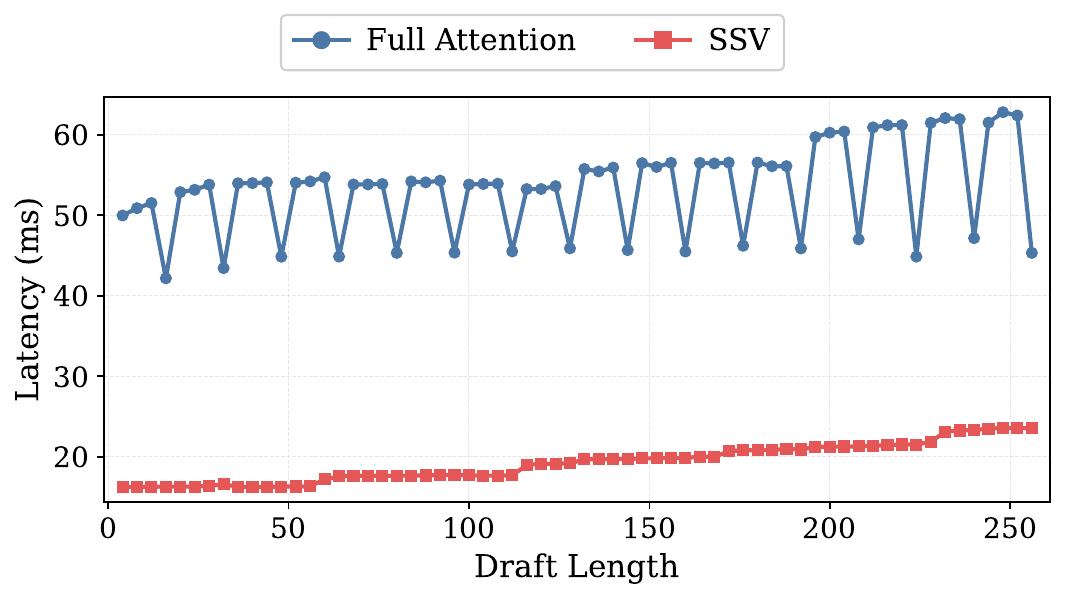}
    \caption{
        Forward latency versus draft length \(\gamma\) on a Llama3-1B backbone at 32K context; \name uses the same backbone with its attention layers replaced by NSA-based sparse verification.
        % The different trends suggest that sparse verification changes how draft length interacts with verifier latency, motivating profile-guided planning over complete sparse-verification strategies.
    }
    \label{fig:fwd_diff_gamma}
    \Description{}
\end{figure}
Compared with full-attention speculative decoding, sparse-attention speculative decoding introduces several changes as follows: 1) \emph{Different draft-length behavior.} As shown in Figure~\ref{fig:fwd_diff_gamma}, full attention and \name variants show different forward-latency trends as the draft length \(\gamma\) changes, so the draft configuration preferred by a dense verifier may not be optimal for sparse verification; 2) \emph{Larger configuration space.} Sparse verification introduces additional choices, including traversal order, coarsening mode, coarsening factor, and refresh/reuse schedule. These choices interact with draft construction because changing \(\gamma\) also changes the realized verifier batch size, the opportunity for cross-query coarsening, and the extent to which sparse-routing and fusion overheads are amortized; and 3) \emph{Prompt-dependent preference.} As motivated by Insight~3 in Section~\ref{sec:key_observations}, different prompts can favor different configurations because acceptance behavior and selected-index overlap are input-dependent.

This section presents \name's profile-guided orchestration policy, using EAGLE-3~\cite{li2025eagle} as the primary framework. We first formalize the joint strategy space and throughput objective, then describe how offline profiles preselect strong initial strategies, and finally introduce a lightweight runtime guard that dynamically refines choices to handle prompt-specific deviations.
% This section presents \name's profile-guided prompt-adaptive orchestration policy. It first defines the joint strategy space and accepted-token-throughput objective, then explains how offline profiles preselect strong strategies, and finally describes how a lightweight runtime guard refines the choice when prompt-specific acceptance deviates from the profile. We use EAGLE-3~\cite{li2025eagle} as the primary framework for integration and exposition, while preserving the raw and compressed KV-cache invariants required by NSA-style sparse attention.

\subsection{Strategy Space and Optimization Problem}
\label{sec:strategy_coupling}

\stitle{Strategy space.}
\name formulates sparse speculative verification as a joint strategy-selection problem. To navigate the interacting draft-side and target-side choices, we categorize the search space into three types of variables, denoted by \(\theta_d\), \(\theta_s\), and \(\mathcal{P}\):
\begin{itemize}[leftmargin=*, topsep=2pt, itemsep=0pt]
    \item \textbf{Full-attention variables $\theta_d=(D, k, \mathcal{T})$:} Draft-tree depth $D$, branching width $k$, and traversal order $\mathcal{T}$. These dictate the number of verifier queries and their positional adjacency in the batch.
    \item \textbf{Sparse-attention variables $\theta_s=(C, M, \mathcal{S})$:} Thread coarsening factor $C$, coarsening mode $M$, and refresh/reuse schedule $\mathcal{S}$. These control how overlapping KV blocks are shared across queries and how often selected indices are recomputed across layers.
    \item \textbf{User-specified constraint $\mathcal{P}$:} A precision class $\mathcal{P}$ that bounds the valid subspace by restricting the strategy to one of four levels: \emph{Strict} (\emph{exact} coarsening, all-refresh schedule), \emph{Reuse-only} (\emph{exact} coarsening, conservative refresh/reuse schedule), \emph{Approx-only} (\emph{approximate} coarsening, all-refresh schedule), or \emph{Approx+Reuse} (both \emph{approximate} coarsening and index reuse). This categorization explicitly balances exact sparse-attention semantics with peak verification efficiency.
\end{itemize}

\stitle{Problem formulation.}
We formulate framework-level planning as a constrained discrete optimization problem. For a context-length regime $r$ and user-specified precision class $\mathcal{P}$, \name selects the strategy tuple $(\theta_d,\theta_s)$ that maximizes the expected accepted-token throughput:
\[
    \begin{aligned}
        \mathop{\arg\max}_{\substack{(\theta_d,\theta_s)}}
        \frac{\mathbb{E}[A(\theta_d,\theta_s \mid r,\mathcal{P})]}
        {\mathbb{E}[T(\theta_d,\theta_s \mid r,\mathcal{P})]}.
    \end{aligned}
\]
where $A(\theta_d,\theta_s \mid r,\mathcal{P})$ is the number of accepted tokens per verification step, and $T(\theta_d,\theta_s \mid r,\mathcal{P})$ is the end-to-end step latency. $\mathbb{E}[\cdot]$ denotes the expectation over prompts and verifier queries under regime $r$ and precision class $\mathcal{P}$.

\subsection{Profile-Guided Strategy Preselection}
\label{sec:planning_policy}

\stitle{Profiling.}
To solve the optimization problem without runtime overhead, \name builds an offline profiler on the target platform. Given a small calibration prompt set, the profiler evaluates the valid candidate strategies for each $(r,\mathcal{P})$ by running the end-to-end speculative decoding workflow. During execution, it explicitly measures the expected accepted tokens $\mathbb{E}[A(\theta_d,\theta_s \mid r,\mathcal{P})]$ and step latency $\mathbb{E}[T(\theta_d,\theta_s \mid r,\mathcal{P})]$ to compute the accepted-token throughput objective.
The resulting profile is organized as a lookup table indexed by the tuple $(r, \mathcal{P})$. Each table entry contains a ranked list of valid strategy pairs $(\theta_d,\theta_s)$ sorted by descending throughput. Alongside the strategy, the table stores the expected acceptance rate $\mathbb{E}[A(\theta_d,\theta_s \mid r,\mathcal{P})]$, which serves as the reference threshold for the runtime guard.

% \wzb{Describe how to profile, what is given as input, what is measured, and how the profile is organized.}
% \name instantiates the planning objective with an offline deployment profile built on the target serving platform. The profiler takes as input a finite set of complete strategy tuples \(x=(D,k,\mathcal{T},C,M,\mathcal{S})\), context-length regimes, a deployment-specified precision class \(\mathcal{P}\), and a small calibration prompt set. For each candidate and regime, it runs the same end-to-end speculative decoding workflow used at runtime and measures accepted-token throughput, step latency, and the average accepted tokens per verification step. The resulting profile is a table indexed by context-length regime and precision class; each entry stores valid strategy tuples ranked by measured accepted-token throughput, together with their throughput and expected accepted-token rate used later by the runtime guard.\zzy{The above is the version after adding information, but the expression still needs to be modified.}

\stitle{Preselection.}
In our implementation, the offline profile forms a compact 192-entry lookup table spanning four context-length buckets (\(0\)--\(4\)K, \(4\)--\(8\)K, \(8\)--\(12\)K, and \(12\)--\(16\)K), four precision classes and 12 ranked candidates in each bucket. At runtime, \name bypasses expensive online grid search via a direct $O(1)$ table lookup. Given the initial context length and user-specified precision class $\mathcal{P}$, it maps them to the regime $r$ and retrieves the highest-ranked valid strategy $(\theta_d^\ast,\theta_s^\ast)$.
This lookup gives a low-overhead initial configuration; since offline profiles rely on small calibration sets and coarse buckets, the runtime guard (Section~\ref{sec:prompt_adaptive_refinement}) later refines the strategy when prompt-specific behavior deviates.
% \wzb{add a paragraph describing why profiling is limited, I think it is benchmark on a small set of prompts, and the best strategy can be prompt-dependent.}\zzy{ok}

% The deployment profile provides a strong initial ranking rather than a per-prompt optimum. It is built from a small calibration prompt set and coarse context-length buckets, while the best strategy can still vary with prompt-specific acceptance behavior and sparse-layout overlap. Therefore, the top-ranked tuple for a bucket and precision class may not be optimal for every new prompt, motivating the prompt-adaptive refinement policy below.

\subsection{Prompt-Adaptive Strategy Refinement}
\label{sec:prompt_adaptive_refinement}

\stitle{Prompt sensitivity.}
A small profiling slice under the same deployment setting shows that the best profiled strategy is not uniform across prompt types. As shown in Table~\ref{tab:prompt_sensitivity}, different prompts select different profiled strategies.
This motivates prompt-aware reselection when observed acceptance deviates from the profile.

% \begin{table}[t]
%     \centering
%     \caption{Prompt sensitivity of the best profiled strategy.}
%     \label{tab:prompt_sensitivity}
%     \small
%     \begin{tabular}{l l c}
%         \hline
%         Prompt      & Best strategy                & Tok/s \\
%         \hline
%         chat-gen.   & \texttt{dfs-med-base-d6k10}  & 233.4 \\
%         chat-plan.  & \texttt{bfs-large-deep-d7k8} & 232.8 \\
%         plain-plan. & \texttt{dfs-small-deep-d7k8} & 225.9 \\
%         code-debug  & \texttt{dfs-small-deep-d7k8} & 266.7 \\
%         \hline
%     \end{tabular}
% \end{table}

\begin{table}[t]
    \centering
    \caption{Prompt sensitivity of profiled strategies in the \(0\)--\(4\)K context bucket under \(\mathcal{P}=\emph{Strict}\). Values report accepted-token throughput in tok/s, with the best result for each prompt shown in bold.} %\zzy{The best result for each prompt is shown in bold.}
    \label{tab:prompt_sensitivity}
    \small
    \setlength{\tabcolsep}{4.2pt}
    \begin{tabular}{@{}lccc@{}}
        \toprule
        \textbf{Prompt}         &
        \textbf{BFS-large-deep} &
        \textbf{DFS-small-deep} &
        \textbf{DFS-med-base}                                                      \\
        \textbf{}               &
        \textbf{(tok/s)}        &
        \textbf{(tok/s)}        &
        \textbf{(tok/s)}                                                           \\
        \midrule
        chat-gen.               & 229.7          & 229.8          & \textbf{233.4} \\
        chat-plan.              & \textbf{232.8} & 222.9          & 225.0          \\
        plain                   & 202.3          & \textbf{225.9} & 210.2          \\
        code                    & 242.2          & \textbf{266.7} & 259.2          \\
        \bottomrule
    \end{tabular}
\end{table}

% \begin{algorithm}[t]
%     \caption{Prompt-aware use of the offline profile. \textsc{Select} returns the best profiled valid strategy for the current regime, precision class, and runtime state \(s\), which records the current profile-selection state maintained by the guard. During the initial speculative-decoding steps, the guard triggers reselection if observed acceptance persistently falls below the profiled expectation.}
%     \label{alg:profile_guided_selection}
%     \begin{algorithmic}[1]
%         \STATE \textbf{Input:} profile $\mathcal{R}$, regime $r$, class $\mathcal{P}$
%         \STATE \textbf{State:} runtime state $s$, guard $G$
%         \STATE $x \gets \textsc{Select}(\mathcal{R}, r, \mathcal{P}, s)$
%         \STATE reset $G$
%         \FOR{each verification step $t$}
%         \STATE Verify with $x$ and observe accepted tokens $A_t$
%         \STATE Update $s$ and $G$ using $A_t$
%         \IF{$G$ triggers reselection}
%         \STATE $r \gets R(L_t)$
%         \STATE $x \gets \textsc{Select}(\mathcal{R}, r, \mathcal{P}, s)$
%         \STATE reset $G$
%         \ENDIF
%         \ENDFOR
%     \end{algorithmic}
% \end{algorithm}

\begin{algorithm}[t]
    \caption{Prompt-adaptive runtime strategy refinement.}
    \label{alg:profile_guided_selection}
    \begin{algorithmic}[1]
        \STATE \textbf{Input:} Profile $\mathcal{R}$, regime $r$, precision class $\mathcal{P}$
        \STATE $(\theta_d,\theta_s) \gets \textsc{Preselect}(\mathcal{R}, r, \mathcal{P})$
        \FOR{each verification step $t$}
        \STATE Run verification, obtain $A_t,T_t$
        % \STATE Verify $A_t,T_t$ with $(\theta_d,\theta_s)$
        \IF{Within the early steps}
        \STATE $(\theta_d,\theta_s) \gets \textsc{Refine}(\mathcal{R}, r, \mathcal{P}, A_t,T_t)$
        \ENDIF
        \ENDFOR
    \end{algorithmic}
\end{algorithm}

\emph{The early decoding window provides enough observations for runtime adaptation.}

\name therefore refines the preselected strategy only during the initial verification steps. At each step, it records the accepted tokens and step latency, then uses these observations to decide whether the profile-selected strategy should be updated.

\stitle{Prompt-aware runtime refinement.}
As outlined in Algorithm~\ref{alg:profile_guided_selection}, \name starts from the profile-preselected strategy and runs verification with the current $(\theta_d,\theta_s)$. Each step produces the accepted-token count $A_t$ and latency $T_t$, which are already available from speculative verification.
During the early steps, \name updates the context-length state as tokens are accepted and invokes \textsc{Refine} with the profile, regime, precision class, and the latest observations. Inside \textsc{Refine}, \name smooths the observed accepted-token counts and compares the smoothed value with the expected acceptance $\mathbb{E}[A(\theta_d,\theta_s \mid r,\mathcal{P})]$ stored for the active strategy. In our experiments, the smoothing coefficient is $\alpha=0.40$; after an 8-step warmup, if the smoothed value remains below $0.85\times$ the profiled expectation for 5 consecutive steps, \textsc{Refine} switches to the next highest-ranked valid strategy in the current profile.
To avoid oscillation, \name permits at most two profile transitions per request and selects the best configuration explored in the current run if the mismatch persists. These constants are fixed across context buckets, precision classes, and held-out prompts. Since this policy relies only on statistics already emitted by verification, its negligible bookkeeping overhead is further evaluated in~\refsec{q4_planning}.

\section{Evaluation}
\label{sec:evaluation}

This section presents a comprehensive evaluation of \name. We aim to investigate the following research questions to validate the effectiveness of our proposed optimizations:

\begin{itemize}[leftmargin=*, topsep=2pt, itemsep=0pt]
  \item \textbf{Q1: Integration with Speculative Frameworks.} When integrated with a real speculative decoding framework such as EAGLE-3~\cite{li2025eagle}, how much end-to-end generation throughput can the full \name system achieve?
  \item \textbf{Q2: Verification-Stage Performance.} What is the impact of \name on speculative verification latency?
  \item \textbf{Q3: Kernel Performance Breakdown.} How do the various implementations of the \name kernel perform? What is the breakdown performance of \name kernels?
  \item \textbf{Q4: Effectiveness of Verification Planning.} Does joint planning over draft construction and sparse verification improve accepted-token throughput over fixed configurations, and can lightweight runtime refinement recover from prompt-specific profile mismatch?
\end{itemize}

\stitle{Evaluation setup.}
Unless otherwise stated, all experiments are conducted on an NVIDIA H100 PCIe GPU~\cite{nvidia_h100}. Our baseline starts from the public NSA Triton reference implementation~\cite{nsa_triton_repo,tillet2019triton} and extends to the verification-oriented execution. We instantiate the standard NSA configuration~\cite{yuan2025native} with compression block size $l = 32$ and stride $d = 16$, selected block size $l' = 64$ and selected block count $n = 16$, and sliding-window size $w = 512$, together with 32 query heads, 8 KV heads, and head dimension 64. For end-to-end experiments, prompts are drawn from Children-Stories-Collection~\cite{ajibawa2023childrenstories}. Unless specified, we use batch size 1 and \texttt{bfloat16} precision~\cite{kalamkar2019study}. All reported latencies are measured with CUDA events via \texttt{torch.cuda.Event}~\cite{pytorch_cuda_event}.

%%%%%%%%%%%%%%%%%%%%%%%%%%%%%%%%%%%%%%%%%%%%%%%%%%%%%
\subsection{End-to-End Evaluation in EAGLE-3 (Q1)}
\label{sec:q1}
%%%%%%%%%%%%%%%%%%%%%%%%%%%%%%%%%%%%%%%%%%%%%%%%%%%%%

\begin{figure*}[t]
  \centering
  \subcaptionbox{Vanilla NSA\label{fig:throughput_comparison_vanilla}}[0.272\textwidth]{%
    \includegraphics[width=\linewidth]{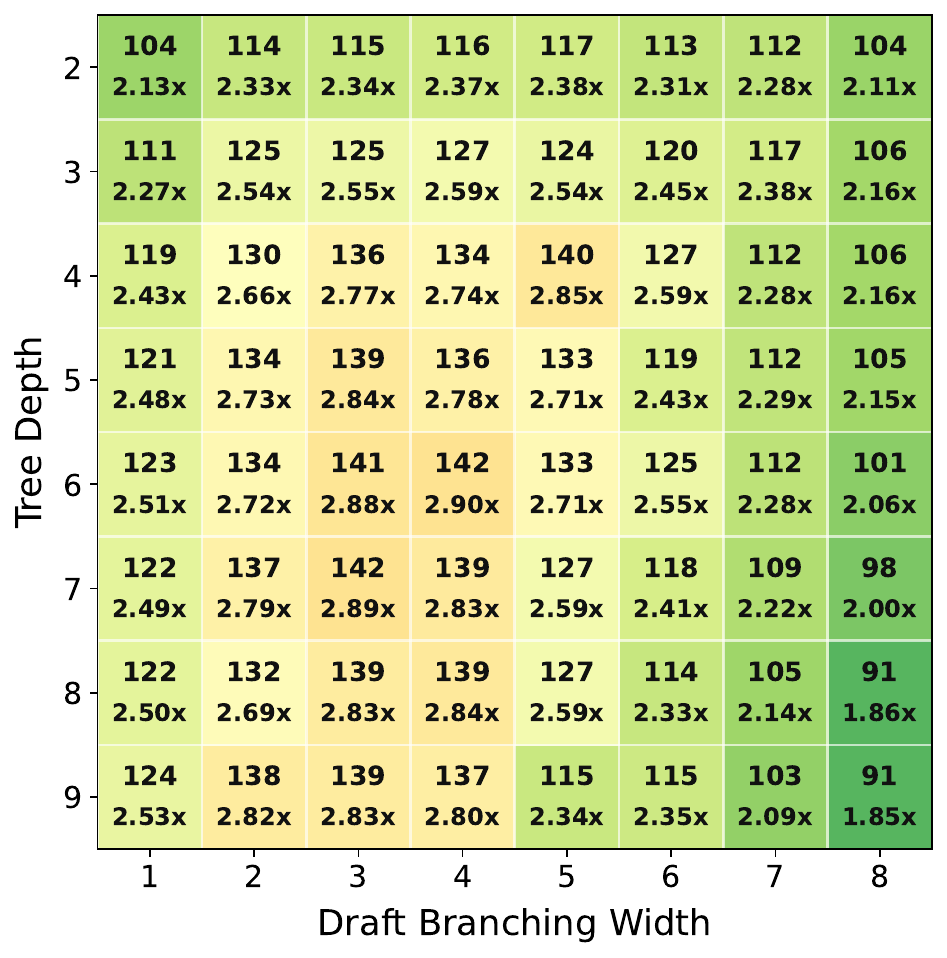}%
  }
  \hspace{0.004\textwidth}
  \subcaptionbox{\name\label{fig:throughput_comparison_specsa}}[0.272\textwidth]{%
    \includegraphics[width=\linewidth]{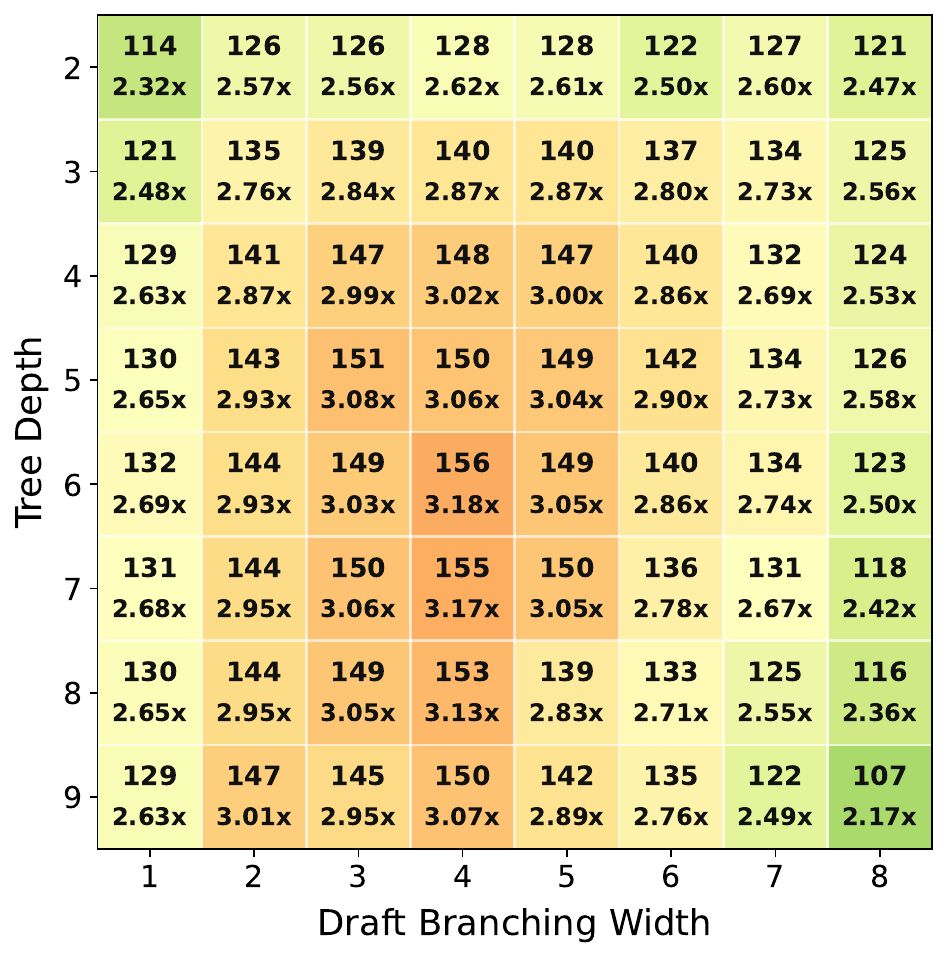}%
  }
  \hspace{0.004\textwidth}
  \subcaptionbox{\name with reuse layers\label{fig:throughput_comparison_specsa_s}}[0.272\textwidth]{%
    \includegraphics[width=\linewidth]{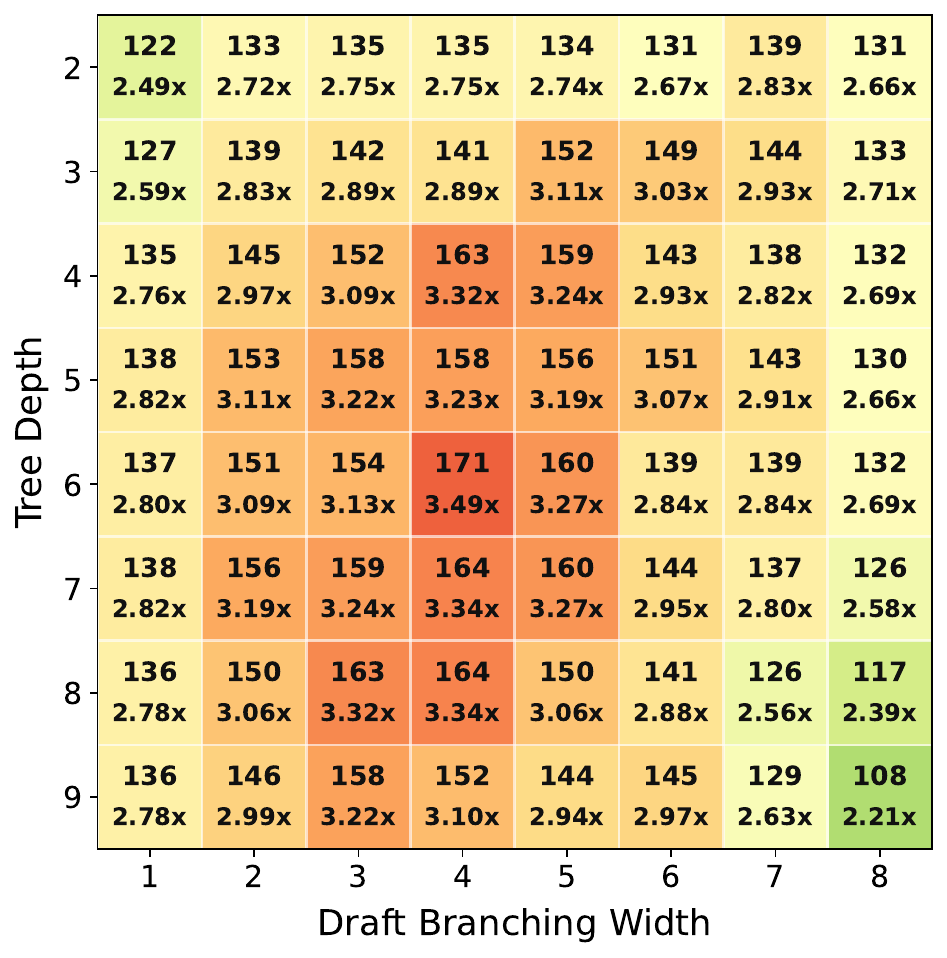}%
  }
  \hspace{0.004\textwidth}
  \includegraphics[width=0.14\textwidth,height=0.21\textheight,keepaspectratio]{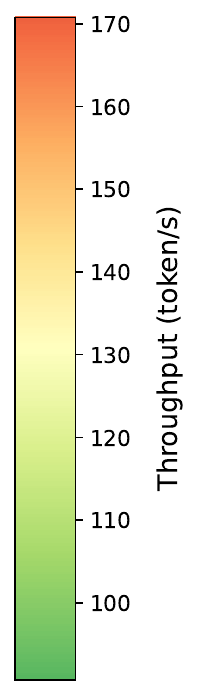}

  \caption{
    End-to-end generation throughput under EAGLE-3 with different draft-tree shapes. Each cell shows throughput on the first line and speedup on the second line, where the speedup baseline is the NSA decode throughput (49 token/s).
  }
  \label{fig:throughput_comparison}
  \Description{}
\end{figure*}

\stitle{Configuration and metric.}
To evaluate whether \name's optimizations translate to visible end-to-end gains, we integrate vanilla NSA~\cite{nsa_triton_repo} and \name into EAGLE-3~\cite{li2025eagle}. We evaluate a 1.2B-parameter NSA target model pretrained on Children-Stories-Collection~\cite{ajibawa2023childrenstories}. All experiments use a 16K context length, temperature 0.0, and 256 generated tokens. Under this setting, the plain autoregressive decode baseline reaches 49 token/s, and serves as the common speedup reference. We compare end-to-end throughput across draft-tree shapes parameterized by tree depth $D$ and branching width $k$. For larger draft lengths ($\gamma \ge 64$), we enable the exact merged-schedule variant with grouping factor $C=2$. In the \name{} with reuse layers setting, the refresh/reuse schedule is calibrated using a training-free IndexCache-style policy~\cite{bai2026indexcache}, yielding \(\mathcal{S} = \{1,2,3,6,8,10,11,15\}\).

\stitle{Results.}
As shown in Figure~\ref{fig:throughput_comparison}, integrating sparse verification with speculative decoding yields large end-to-end gains over the 49 token/s decode baseline. Across the full grid of evaluated $(D,k)$ settings, vanilla NSA integrated with EAGLE-3 reaches 91--142 token/s, corresponding to $1.85\times$--$2.90\times$ speedup. \name consistently improves throughput across these tree shapes: without reuse layers, it reaches 107--156 token/s, or $2.17\times$--$3.18\times$ speedup, and enabling reuse layers further boosts throughput to 108--171 token/s, or $2.21\times$--$3.49\times$ speedup.

The highest absolute throughput is achieved by \name with reuse layers at $(D=6,k=4)$, where it reaches 171 token/s, or $3.49\times$ over the decode baseline. Overall, these results show that the combination of sparse attention with speculative decoding delivers substantial end-to-end acceleration, and that the additional optimizations in \name, especially reuse-layer execution, can further improve throughput across the evaluated draft-tree settings.

%%%%%%%%%%%%%%%%%%%%%%%%%%%%%%%%%%%%%%%%%%%%%%%%%%%%%
\subsection{Verification-Stage Performance (Q2)}
%%%%%%%%%%%%%%%%%%%%%%%%%%%%%%%%%%%%%%%%%%%%%%%%%%%%%

\begin{figure}[h]
  \centering
  \includegraphics[width=0.95\linewidth]{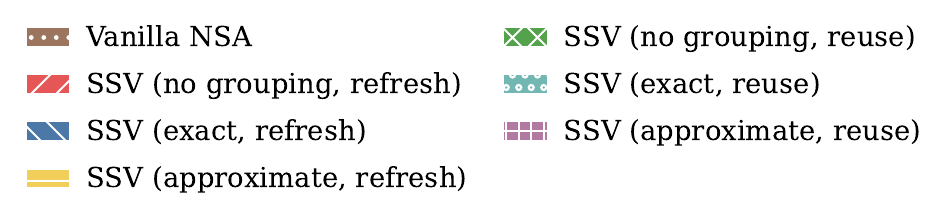}

  \vspace{0.4em}

  \subcaptionbox{Llama3-1B, $\gamma = 4$\label{fig:forward_benchmark_1b_gamma4}}[0.48\linewidth]{%
    \includegraphics[width=\linewidth]{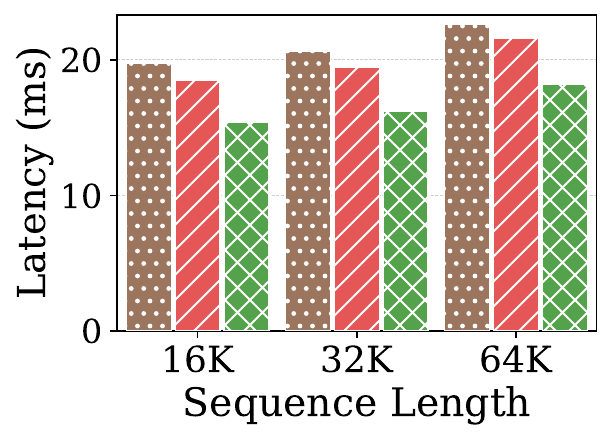}%
  }
  \hfill
  \subcaptionbox{Llama3-8B, $\gamma = 4$\label{fig:forward_benchmark_8b_gamma4}}[0.48\linewidth]{%
    \includegraphics[width=\linewidth]{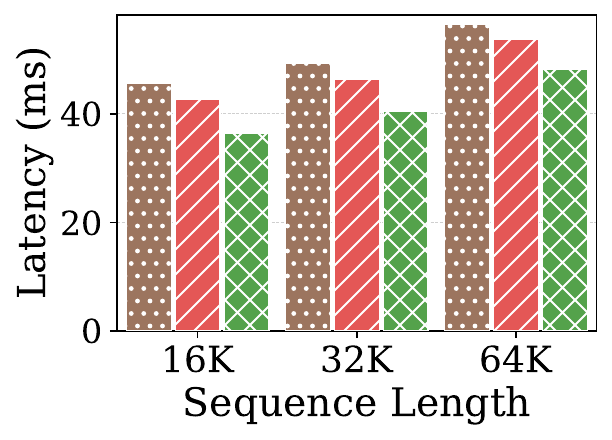}%
  }

  \vspace{0.35em}

  \subcaptionbox{Llama3-1B, $\gamma = 64$\label{fig:forward_benchmark_1b_gamma64}}[0.48\linewidth]{%
    \includegraphics[width=\linewidth]{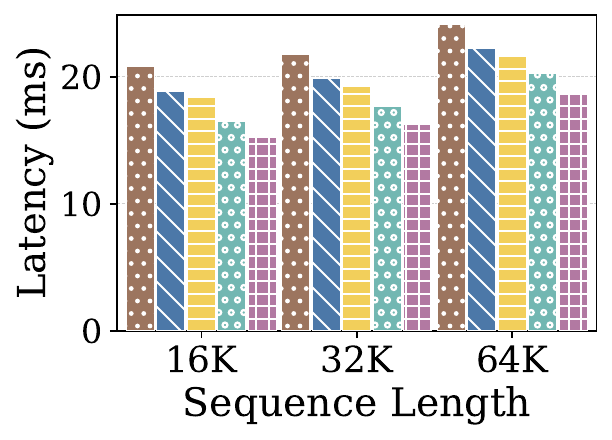}%
  }
  \hfill
  \subcaptionbox{Llama3-8B, $\gamma = 64$\label{fig:forward_benchmark_8b_gamma64}}[0.48\linewidth]{%
    \includegraphics[width=\linewidth]{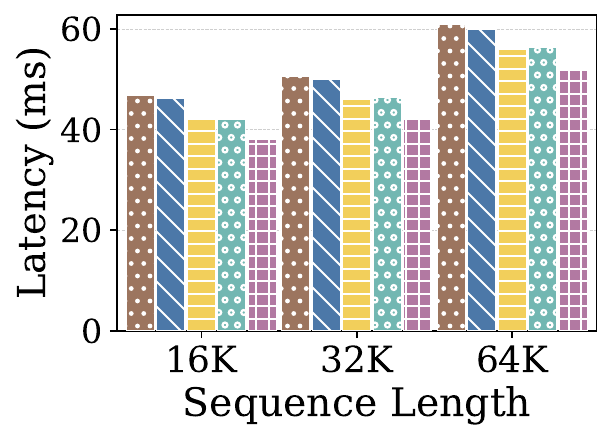}%
  }

  \vspace{0.35em}

  \subcaptionbox{Llama3-1B, $\gamma = 128$\label{fig:forward_benchmark_1b_gamma128}}[0.48\linewidth]{%
    \includegraphics[width=\linewidth]{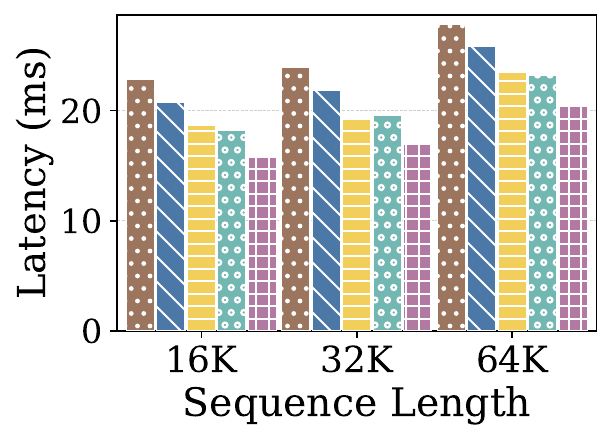}%
  }
  \hfill
  \subcaptionbox{Llama3-8B, $\gamma = 128$\label{fig:forward_benchmark_8b_gamma128}}[0.48\linewidth]{%
    \includegraphics[width=\linewidth]{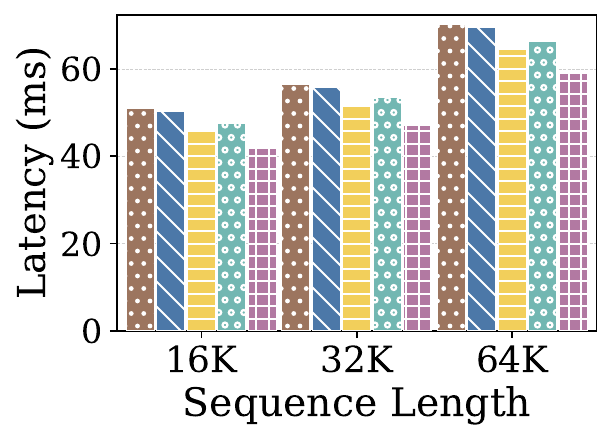}%
  }
  \caption{
    Speculative verification latency on NSA target models built from Llama3-1B and Llama3-8B backbones under different draft lengths $\gamma$ and sequence lengths $N$.
    We compare vanilla NSA~\cite{nsa_triton_repo}, \name without reuse layers, and \name with reuse layers; for $\gamma \in \{64,128\}$, we include the exact merged-schedule variant with $C=2$ and the approximate shared-index variant with $C=4$. Refresh/reuse indicates the layer type, and no grouping/exact/approximate indicates the grouped-query kernel mode.
  }
  \label{fig:forward_benchmark}
  \Description{}
\end{figure}

\stitle{Configuration and metric.}
To benchmark verification latency (Q2), we evaluate NSA target models adapted from Llama3-1B and Llama3-8B backbones~\cite{grattafiori2024llama3,yuan2025native}. We test sequence lengths $N \in \{16\text{K}, 32\text{K}, 64\text{K}\}$ and draft lengths $\gamma \in \{4, 64, 128\}$. For reuse layers, we apply an alternating schedule \(\mathcal{S} = \{1,3,5,\ldots\}\). At larger draft lengths ($\gamma \in \{64,128\}$), we also evaluate our most efficient grouped-query variants: exact merged-schedule ($C=2$) and approximate shared-index ($C=4$). We report verification latency and speedup over vanilla NSA~\cite{nsa_triton_repo}.

\stitle{Results.}
As shown in Figure~\ref{fig:forward_benchmark}, \name consistently accelerates verification, with performance gains stacking clearly as optimizations are combined.
At a short draft length ($\gamma=4$), \name without reuse layers yields modest $1.05\times$--$1.07\times$ speedups over vanilla NSA~\cite{nsa_triton_repo} on both models. Enabling reuse layers significantly boosts this to $1.24\times$--$1.28\times$ (1B) and $1.17\times$--$1.25\times$ (8B). For larger draft lengths ($\gamma \ge 64$), grouped-query execution brings additional independent gains. For instance, on the 1B model, the exact variant ($C=2$) and approximate variant ($C=4$) reach up to $1.10\times$ and $1.23\times$ speedups, respectively. The 8B model follows the same trend, albeit with slightly more modest gains, up to $1.02\times$ and $1.12\times$.

Crucially, the synergy of multiple optimizations delivers the strongest overall results, especially reuse layers plus the approximate shared-index variant. While combining reuse layers with the exact merged-schedule variant ($C=2$) yields substantial speedups of $1.19\times$--$1.26\times$ (1B) and $1.06\times$--$1.12\times$ (8B), upgrading to the approximate shared-index variant ($C=4$) unlocks peak performance across the board. This aggressive combination achieves the highest overall speedups of $1.30\times$--$1.45\times$ (1B) and $1.18\times$--$1.23\times$ (8B). These results confirm that while individual mechanisms provide measurable benefits, their synergy is required to unlock the largest practical acceleration for NSA verification.
\subsection{\name Kernel Benchmarking Results (Q3)}
%%%%%%%%%%%%%%%%%%%%%%%%%%%%%%%%%%%%%%%%%%%%%%%%%%%%%

\begin{figure}[h]
  \centering
  \includegraphics[width=0.95\linewidth]{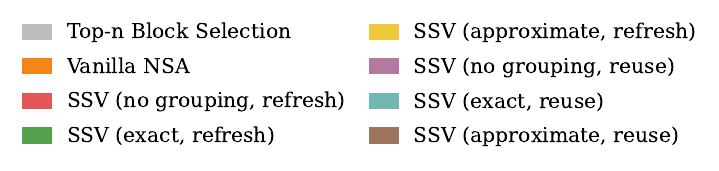}

  \vspace{0.4em}

  \subcaptionbox{$\gamma = 4$, $s = 3$\label{fig:specnsa_kernels_gamma4_s3}}[0.48\linewidth]{%
    \includegraphics[width=\linewidth]{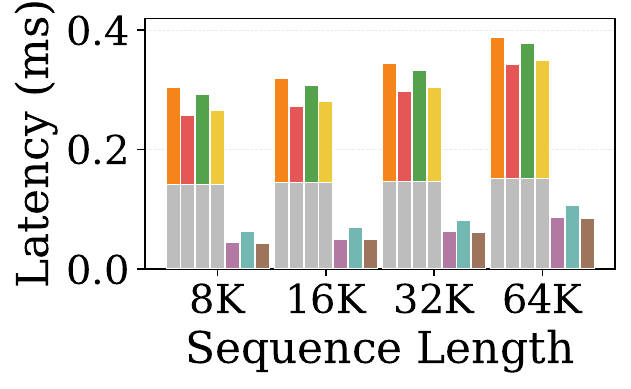}%
  }
  \hfill
  \subcaptionbox{$\gamma = 64$, $s = 3$\label{fig:specnsa_kernels_gamma64_s3}}[0.48\linewidth]{%
    \includegraphics[width=\linewidth]{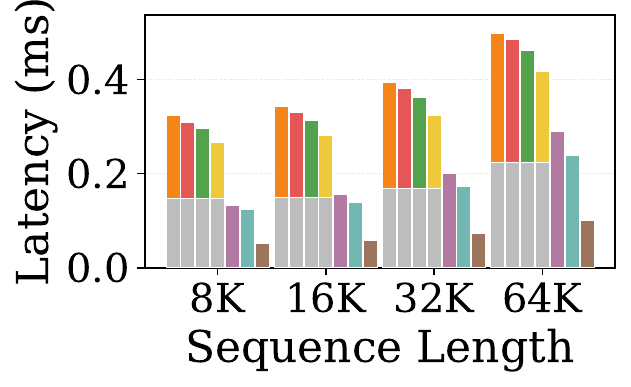}%
  }

  \vspace{0.35em}

  \subcaptionbox{$\gamma = 4$, $s = 6$\label{fig:specnsa_kernels_gamma4_s6}}[0.48\linewidth]{%
    \includegraphics[width=\linewidth]{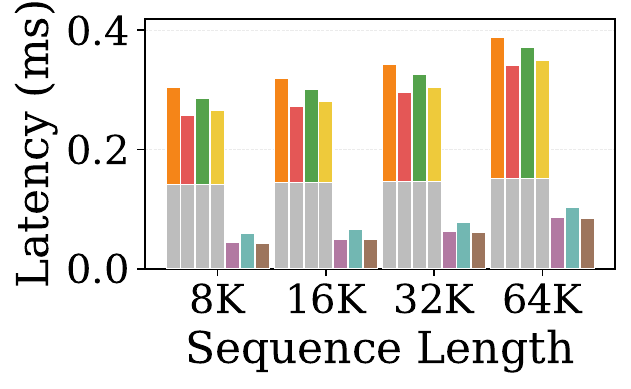}%
  }
  \hfill
  \subcaptionbox{$\gamma = 64$, $s = 6$\label{fig:specnsa_kernels_gamma64_s6}}[0.48\linewidth]{%
    \includegraphics[width=\linewidth]{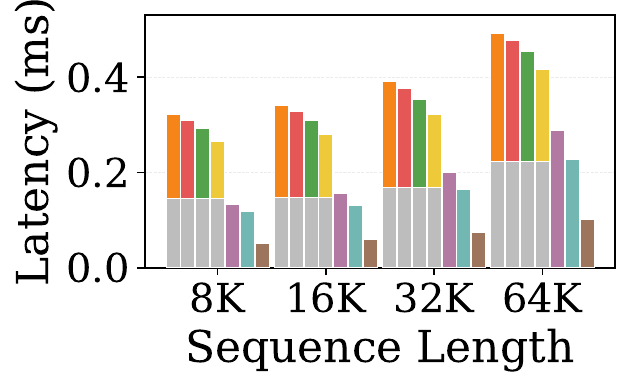}%
  }

  \vspace{0.35em}

  \subcaptionbox{$\gamma = 4$, $s = 10$\label{fig:specnsa_kernels_gamma4_s10}}[0.48\linewidth]{%
    \includegraphics[width=\linewidth]{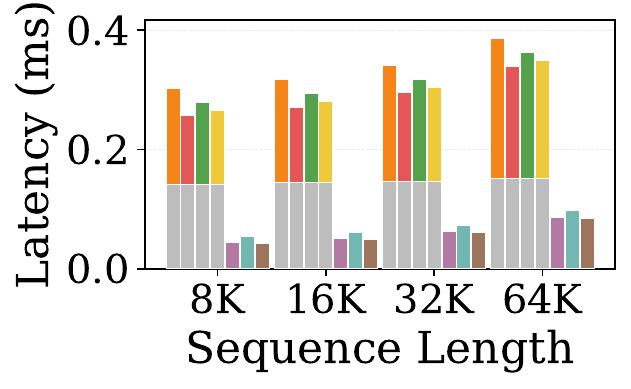}%
  }
  \hfill
  \subcaptionbox{$\gamma = 64$, $s = 10$\label{fig:specnsa_kernels_gamma64_s10}}[0.48\linewidth]{%
    \includegraphics[width=\linewidth]{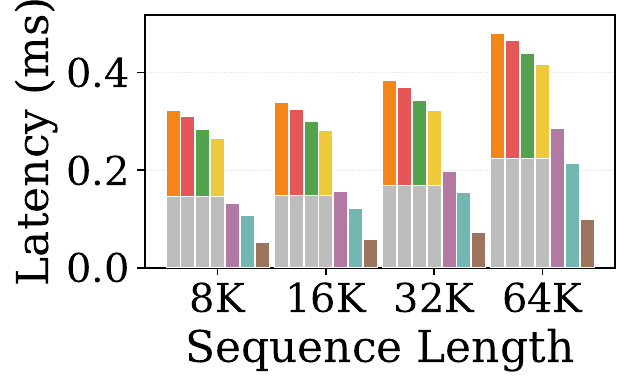}%
  }
  \caption{
    Performance breakdown and ablation of \name kernel variants.
    The charts compare grouped-query kernel variants, cross-query overlap $s$ (the number of shared selected blocks between adjacent queries), and reuse-layer execution under small and large draft lengths ($\gamma = 4$ and $\gamma = 64$).
    Refresh/reuse indicates the layer type, and no grouping/exact/approximate indicates the grouped-query kernel mode.
  }
  \label{fig:specnsa_kernels}
  \Description{
    A set of bar charts visualizing the latency of different \name implementation variants.
    The breakdown reveals the substantial overhead of index construction compared to sparse attention computation.
    It also illustrates the transition where grouped-query execution becomes beneficial as $\gamma$ and $s$ increase, and shows the consistent dominance of the fully fused reuse-layer kernels across all settings.
  }
\end{figure}

\stitle{Configuration and metric.}
To benchmark verification kernels (Q3), we evaluate draft lengths $\gamma \in \{4, 64\}$ across sequence lengths $N \in \{8\text{K}, 16\text{K}, 32\text{K}, 64\text{K}\}$. We fix the selected-block count $n = 16$ and control cross-query redundancy by varying the adjacent-query overlap $s = |I_t^{(j)} \cap I_{t-1}^{(j)}| \in \{3, 6, 10\}$. This range explicitly covers both the theoretical lower bound ($s=3$, due to mandatory initial/local blocks~\cite{yuan2025native}) and representative high-overlap scenarios ($s \in \{6, 10\}$) observed in real EAGLE traces. We report speedups over the vanilla NSA baseline~\cite{nsa_triton_repo} in Fig.~\ref{fig:specnsa_kernels}.
% For Q3, we benchmark isolated verification kernels under draft lengths $\gamma \in \{4,\, 64\}$ and sequence lengths $N \in \{8\text{K},\, 16\text{K},\, 32\text{K},\, 64\text{K}\}$. To control cross-query redundancy, we fix $n = 16$ and vary the adjacent-query overlap as $s \in \{3,\, 6,\, 10\}$, where the cross-query overlap $s = |I_t^{(j)} \cap I_{t-1}^{(j)}|$ denotes the number of shared selected blocks between adjacent queries. Here, $s=3$ is the lower bound under the standard NSA configuration~\cite{yuan2025native}, because the selected-block count $n=16$ includes one fixed initial block and two local blocks that are always activated. In real EAGLE verification traces, overlap is typically above this lower bound, and values around $s \in \{6,10\}$ are commonly observed. We therefore use $s=\{3,6,10\}$ to cover the lower bound and representative common-overlap settings. We report kernel latency and speedup over the vanilla NSA baseline~\cite{nsa_triton_repo}.

\stitle{Comparison with Vanilla NSA.}
For short drafts ($\gamma = 4$), the primary gains stem from kernel fusion. Compared to the vanilla NSA baseline~\cite{nsa_triton_repo}, \name without reuse layers achieves a $1.14\times$--$1.18\times$ speedup. Enabling full fusion in reuse layers removes significant index-construction overhead, catapulting the speedup to $4.45\times$--$6.86\times$.

For longer drafts ($\gamma = 64$), grouped-query execution introduces additional independent gains. On their own, the exact ($C=2$) and approximate ($C=4$) variants yield speedups up to $1.13\times$ and $1.22\times$, respectively. However, the most profound acceleration occurs when these are combined with reuse layers. While \name with reuse layers alone achieves up to $2.44\times$, pairing it with the exact variant reaches $2.09\times$--$2.99\times$, and combining it with the approximate variant yields a $4.81\times$--$6.30\times$ speedup. Overall, while grouping provides measurable improvements, eliminating repeated index construction remains the dominant source of acceleration.

\stitle{Comparison between variants of the \name kernel.}
The benefits of grouped-query execution are highly dependent on draft length, scaling effectively at larger $\gamma$ (e.g., $\gamma = 64$). Furthermore, as cross-query overlap $s$ increases from $3$ to $10$, the grouped variants yield increasingly higher gains. This confirms that larger verification batches provide sufficient adjacent queries to successfully amortize scheduling and KV-loading costs.

Crucially, cross-layer reuse and cross-query grouping act as highly complementary mechanisms. Skipping $I_t^{(j)}$ construction enables deeper kernel fusion, allowing reuse paths to consistently dominate non-reuse baselines. Meanwhile, the higher $s$ overlap in large $\gamma$ settings maximizes the efficiency of joint query execution. Together, they eliminate the index bottleneck and expose a shorter fused path.

\stitle{Performance breakdown of \name.}
Constructing selected block indices is a severe memory-bound bottleneck in NSA-style verification, accounting for $45\%$--$56\%$ of the total kernel runtime. In small-$\gamma$ scenarios ($\gamma=4$), this index-selection cost even exceeds the sparse attention computation itself (averaging $1.05\times$ the attention cost), and it remains substantial ($0.84\times$--$0.88\times$) at $\gamma=64$.

This breakdown highlights exactly why reuse layers are so transformative. By inheriting $I_t^{(j)}$ from a preceding refresh layer instead of recomputing it, \name entirely bypasses this massive overhead and unlocks more aggressive kernel fusion. Therefore, the reuse-layer design is not a marginal tweak, but the fundamental mechanism translating NSA's algorithmic sparsity into concrete hardware acceleration.
\subsection{Throughput-Aware Verification Planning (Q4)}
\label{sec:q4_planning}
%%%%%%%%%%%%%%%%%%%%%%%%%%%%%%%%%%%%%%%%%%%%%%%%%%%%%

\stitle{Configuration and metric.}
For Q4, we evaluate the planner from Section~\ref{sec:planner} on held-out prompts using accepted-token throughput. We compare three settings: Base, Static-best, and Best+R. Base is a non-adaptive EAGLE-3 configuration with BFS traversal, a 128-token draft-tree budget, depth \(D=6\), draft TopK \(k=10\), exact coarsening with \(C=2\), and an all-\texttt{F} sparse-verification schedule. Static-best uses the top-ranked profiled strategy for each context bucket and precision class without runtime refinement, while Best+R additionally enables runtime refinement. All three settings generate 128 output tokens with temperature set to 0.0, and run with a maximum context length of 16K. Thus, planner gains are measured against a fixed strategy under the same model, workload, and measurement protocol. The offline profile covers 16 bucket-class pairs, with 12 ranked candidate strategies stored for each pair. Static-best uses the top-ranked candidate for the active pair, while Best+R starts from the same candidate and may refine the choice using runtime observations. Table~\ref{tab:planner} therefore reports one row for each bucket-class pair.

\begin{table}[!htbp]
  \centering
  \footnotesize
  \setlength{\tabcolsep}{4.5pt}
  \caption{Effectiveness of throughput-aware planning on held-out prompts. Static-best uses the top-ranked profiled strategy without runtime refinement, Best+R adds runtime refinement, Gain reports the improvement of Best+R over Base, and RR indicates whether refinement is triggered. Throughput is measured in accepted tokens/s.}
  \label{tab:planner}
  \begin{tabular}{@{}llrrrrc@{}}
    \toprule
    \textbf{Bucket} & \textbf{Class} & \textbf{Base} & \textbf{Static-best} & \textbf{Best+R} & \textbf{Gain} & \textbf{RR} \\
    \midrule
    0--4K           & Strict         & 196.3         & 164.2                & 192.4           & $-2.0\%$      & yes         \\
    0--4K           & Reuse-only     & 196.3         & 232.7                & 230.7           & $17.5\%$      & no          \\
    0--4K           & Approx-only    & 196.3         & 207.6                & 206.8           & $5.3\%$       & no          \\
    0--4K           & Approx+Reuse   & 196.3         & 206.1                & 247.9           & $26.3\%$      & yes         \\
    4--8K           & Strict         & 169.0         & 197.0                & 200.5           & $18.6\%$      & no          \\
    4--8K           & Reuse-only     & 169.0         & 216.0                & 217.4           & $28.6\%$      & no          \\
    4--8K           & Approx-only    & 169.0         & 182.7                & 181.8           & $7.6\%$       & no          \\
    4--8K           & Approx+Reuse   & 169.0         & 192.4                & 192.4           & $13.8\%$      & no          \\
    8--12K          & Strict         & 168.6         & 171.5                & 172.3           & $2.2\%$       & no          \\
    8--12K          & Reuse-only     & 168.6         & 189.3                & 189.4           & $12.3\%$      & no          \\
    8--12K          & Approx-only    & 168.6         & 176.6                & 176.5           & $4.7\%$       & no          \\
    8--12K          & Approx+Reuse   & 168.6         & 192.0                & 192.5           & $14.2\%$      & no          \\
    12--16K         & Strict         & 137.4         & 156.2                & 154.4           & $12.4\%$      & no          \\
    12--16K         & Reuse-only     & 137.4         & 165.5                & 164.7           & $19.9\%$      & no          \\
    12--16K         & Approx-only    & 137.4         & 137.2                & 168.0           & $22.3\%$      & yes         \\
    12--16K         & Approx+Reuse   & 137.4         & 182.6                & 183.0           & $33.2\%$      & no          \\
    \bottomrule
  \end{tabular}
\end{table}

\stitle{Offline ranking.}
The offline profiles show that no single parameter determines the best strategy: top choices vary across buckets and precision classes, with profiling-set gains ranging from 4.0\% to 33.5\% over the fixed baseline. On held-out prompts, Static-best improves throughput by 10.6\% over Base on average, indicating that complete strategy tuples transfer beyond the profiling set. The remaining variation is prompt-specific: some rows, such as 0--4K Strict, expose a mismatch between the profiled top choice and the held-out prompt behavior, while small-gain rows such as 8--12K Strict and 8--12K Approx-only occur because the fixed baseline is already strong. These results support the need for runtime refinement on top of offline ranking.

\stitle{Runtime refinement.}
Best+R improves average throughput by 14.4\% over Base and by 3.4\% over Static-best. Runtime refinement is triggered in 3 of the 16 bucket-class settings, targeting cases where the initial profiled choice mismatches the held-out prompt behavior.
The 0--4K Strict row illustrates this role: Static-best drops from 196.3 to 164.2 tok/s, but runtime refinement brings throughput back near Base at 192.4 tok/s. In 12--16K Approx-only, the same mechanism turns near-Base Static-best performance into a 22.3\% gain. Overall, these results show that lightweight acceptance validation can correct prompt-specific profile mismatches without turning planning into online search.

\stitle{Refinement validation.}
We validate runtime refinement on a deployment profile that selects one strategy per context bucket: Approx+Reuse for 0--4K, Reuse-only for 4--8K, Approx+Reuse for 8--12K, and Approx+Reuse for 12--16K. With strategy refinement disabled, the bookkeeping needed for acceptance validation changes throughput by only \(-1.1\%\) to \(+2.5\%\) across buckets, which is within the run-to-run variation of our decode benchmark. Table~\ref{tab:guard_sensitivity} reports a one-factor sweep over the refinement constants. We denote the guard constants as the EMA coefficient \(\alpha\), the acceptance-drop ratio \(\rho\), the minimum observation count \(m\), and the hysteresis window \(h\). The default setting uses \(\alpha=0.40\), \(\rho=0.85\), \(m=8\), and \(h=5\). Varying the EMA coefficient \(\alpha\), acceptance-drop ratio \(\rho\), and minimum observation count \(m\) keeps gains in the \(25.2\%\)--\(26.6\%\) range. The main sensitivity is the hysteresis window \(h\): \(h=3\) is too aggressive and triggers six refinement events across the four bucket runs, reducing throughput, while \(h=8\) is too conservative and triggers no refinement events, missing useful corrections. The default \(h=5\) triggers one refinement event and achieves the overall tradeoff.

\begin{table}[t]
  \centering
  \footnotesize
  \caption{Refinement sensitivity on the deployment profile used for validation. Refinement Events counts total refinement events across four bucket runs; each run allows at most two refinements.}
  \label{tab:guard_sensitivity}
  \begin{tabular}{@{}lrrc@{}}
    \toprule
    \textbf{Setting} & \textbf{Throughput} & \textbf{Gain/Base} & \textbf{Refinement Events} \\
    \midrule
    Default          & 209.7               & $+26.6\%$          & 1                          \\
    \(\alpha=0.20\)  & 207.3               & $+25.2\%$          & 2                          \\
    \(\alpha=0.60\)  & 209.0               & $+26.1\%$          & 1                          \\
    \(\rho=0.80\)    & 209.6               & $+26.6\%$          & 1                          \\
    \(\rho=0.90\)    & 207.8               & $+25.4\%$          & 1                          \\
    \(m=4\)          & 209.5               & $+26.4\%$          & 1                          \\
    \(m=16\)         & 209.0               & $+26.1\%$          & 1                          \\
    \(h=3\)          & 197.9               & $+18.8\%$          & 6                          \\
    \(h=8\)          & 199.1               & $+21.0\%$          & 0                          \\
    \bottomrule
  \end{tabular}
\end{table}

%%%%%%%%%%%%%%%%%%%%%%%%%%%%%%%%%%%%%%%%%%%%%%%%%%%%%

\section{Conclusion}

In this paper, we present \name, a sparse verification framework to bridge the gap between dynamic sparse attention and speculative decoding.
\name uses cross-query coarsening and verification-oriented kernel fusion to reduce duplicated KV-block loads and fragmented branch execution, while profile-guided planning adapts the draft-verification strategy to input-dependent acceptance and kernel behavior.
Experiments on H100 GPUs with Llama-based NSA models show that \name substantially improves end-to-end generation throughput, achieving up to $3.49\times$ speedup in our EAGLE-3 integration study.

%%
%% The acknowledgments section is defined using the "acks" environment
%% (and NOT an unnumbered section). This ensures the proper
%% identification of the section in the article metadata, and the
%% consistent spelling of the heading.
% \begin{acks}

% \end{acks}

%%
%% The next two lines define the bibliography style to be used, and
%% the bibliography file.
\bibliographystyle{ACM-Reference-Format}
\bibliography{reference}

\end{document}